\documentclass[12pt]{iopart}

\usepackage{setspace}
\usepackage{iopams}
\usepackage{graphicx,psfrag}
\usepackage{amssymb}
\usepackage{soul}
\usepackage{subfigure}
\usepackage{wasysym}
\usepackage{textcomp}
\usepackage{setspace}
\usepackage{esint}
\usepackage{cite}
\usepackage{epsfig}
\usepackage{textcomp}
\usepackage[dvips,usenames]{color}
\definecolor{violet}{rgb}{0.734,0.558,0.558}
\usepackage{fancyhdr}
\begin{document}

\fancyhf{}
\fancyhead[L]{\textit{\nouppercase{Order-disorder transitions in reactive lattice gases}}}
\fancyhead[R]{\nouppercase{M. Dudka  \textit{\nouppercase{et al}}}}
\fancyfoot[C]{\thepage}

\title[Order-disorder transitions in  lattice gases with annealed reactive constraints]{Order-disorder transitions in  lattice gases with annealed reactive constraints}

\author{Maxym Dudka$^{1,2}$, Olivier B\'enichou$^3$ and Gleb Oshanin$^{3,4}$}

\address{$^1$ Institute for Condensed Matter Physics of the National Academy of Sciences of Ukraine, 1 Svientsitskii st., 79011 Lviv, Ukraine}
\address{$^2$ ~${\mathbb L}^4$ Collaboration \& Doctoral College for the Statistical Physics of Complex Systems, Leipzig-Lorraine-Lviv-Coventry, Europe}
\address{$^3$ Laboratoire de Physique Th\'{e}orique de la Mati\`{e}re Condens\'{e}e, Sorbonne Universit\'es, UPMC Univ Paris 06,  Paris, France}
\address{$^4$ Max Planck Institute for Intelligent Systems, Heisenbergstr. 3,
D-70569 Stuttgart, Germany}
\eads{maxdudka@icmp.lviv.ua, benichou@lptmc.jussieu.fr, oshanin@lptmc.jussieu.fr}

\begin{abstract}
We study equilibrium
properties of catalytically-activated
$A + A \to \oslash$ reactions
 taking place
on a
lattice of adsorption sites.
The particles
undergo continuous exchanges with
a reservoir maintained at a constant
chemical potential $\mu$ and react when they
appear at the neighbouring sites, provided that some reactive conditions are fulfilled.
We model the latter in two different ways:
In the Model I
some fraction $p$ of the {\em bonds}  connecting neighbouring sites possesses special catalytic properties
such that any two $A$s appearing on the sites connected by such a bond instantaneously react and desorb.
In the Model II some fraction $p$ of the adsorption {\em sites} possesses such properties and neighbouring particles react
if at least one of them resides on a catalytic site.
For the case of \textit{annealed} disorder in the distribution of the catalyst,
which is tantamount to the situation when the reaction may take place at any point on the lattice but happens with a finite probability $p$, we provide an exact solution for both models for the interior of an infinitely large Cayley tree - the so-called Bethe lattice.
We show that both models exhibit a rich critical behaviour:
For the annealed Model I it is characterised by a transition into an ordered state
and a re-entrant transition into a disordered phase, which both are continuous. For the annealed Model II,
which represents a rather exotic model of statistical mechanics in which interactions of any particle with its environment have a peculiar Boolean form,
 the transition to an ordered state is always continuous, while the re-entrant transition into the disordered phase may be either continuous or discontinuous, depending on the value of $p$.
\end{abstract}

\vspace{2pc}


\vspace{2pc}


\submitto{JSTAT}

\maketitle


\section{Introduction}

Catalytically activated reactions
involve particles that react only in the presence of another agent - a catalyst - and remain chemically
inactive otherwise. Such processes are widespread in nature and are also involved in a variety of technological and
industrial applications \cite{bond,avnir}. Different kinetic, equilibrium and out-of-equilibrium
properties of such reactions, as well as various theoretical concepts, available analytical and numerical approaches
have been comprehensively reviewed in Ref. \cite{evans}.

The classical textbook approach to the kinetics of such processes focuses only on the average concentrations of the species involved. In consequence, one usually writes systems of
differential rate equations of varying complexity with multiple parameters, as prescribed by the formal-kinetic ``law of mass action'' (see, e.g., Ref. \cite{evans}).
Ziff, Gulari and Barshad (ZGB) \cite{ziff} and subsequently, Ziff and Fichthorn \cite{ziff2}, Fichthorn, Gulari and Ziff \cite{fichthorn},
were apparently first to realise that fluctuations in adsorption/desorption events, fluctuations of coverages and spatial correlations can cause a severe departure from the deterministic descriptions based on the rate equations approach.
Using computer simulations ZGB studied the so-called monomer-dimer model, introduced as an
idealised description of the important process of ${\rm CO}$ oxidation on a catalytic surface, which
revealed an emerging spectacular cooperative behaviour: it was found that the monolayers formed by the
molecules adsorbed on the surface may undergo
discontinuous or continuous  phase transitions  in different parameter ranges.

More specifically, ZGB considered a reaction occurring by the following three steps: a) an irreversible adsorption of ${\rm CO}$ molecules
from their reservoir onto a single-crystal catalyst surface, modelled as a simple square lattice of adsorption sites, b) an irreversible adsorption of ${\rm O_2}$ molecules onto the lattice from their reservoir with the subsequent dissociation of ${\rm O}_2$ into two ${\rm O}$ atoms, each residing on a separate lattice site (while ${\rm CO}$ requires only a single site, which explains why
the model is called the "monomer-dimer" one) and c) an instantaneous (perfect) irreversible reaction between neighbouring absorbed ${\rm CO}$ and ${\rm O}$, followed by an immediate desorption of the reaction product ${\rm CO_2}$, which does not interact further with the system.
It was realised that on a two-dimensional square lattice upon lowering the ${\rm CO}$ adsorption rate the system undergoes a first-order (discontinuous) phase transition from a ${\rm CO}$ saturated inactive phase into a reactive steady state, followed by a continuous transition into an ${\rm O_2}$-saturated inactive phase, which belongs to the same universality class as directed percolation and the Reggeon field theory \cite{5}.
 While such a model evidently discards many important features of real physical systems, such as, e.g., molecular diffusion on the surface, reversibility of the  adsorption of the species involved, imperfect reaction, when two neighbouring adsorbed reactants react only with a finite probability,  and etc., it does reproduce two types of phase transitions which occur in realistic systems \cite{evans}.

A more simple, monomer-monomer model was introduced by Ziff and Fichthorn \cite{ziff2} and was subsequently studied in Refs. \cite{doug,alb,redner}. In this model one focuses on a general reaction scheme $A + B \to \oslash$ where $A$ and $B$ are
two chemically different species but both are monomers, in the sense that both require just a single lattice site for an irreversible adsorption.
It was realised that the monomer-monomer model exhibits a first-order transition from a phase saturated with one of the species to the phase saturated with the other one. Allowing for an adsorption of one of the species, i.e., connecting the system to a reservoir, leads to a continuous transition which also belongs to the directed percolation universality class \cite{redner}.
We note also that
 the fluctuation-induced behaviour of the catalytically-activated
reactions between molecules with more complicated structures has been studied
\cite{alb2,alb3,avr}
and also the influence of  the lateral diffusion of the adsorbed species
on the reaction kinetics (see, e.g. Refs. \cite{krap,marro,osh,osh2,thiago} and references therein) as well as of the
 heterogeneity in the spatial distribution of the catalyst (see, e.g., Refs. \cite{blumen,dietrich}) have been addressed.

As long as one is interested only in
the thermodynamic equilibrium
properties of the monolayers of the adsorbed molecules,
formed in the course of such catalytically-activated reactions,
one may realise that the latter should be
very
similar to the analogous properties of the adsorbates emerging in the
models of  hard-core lattice gases with athermal interactions, in which one studies a reversible
deposition of "hard", non-overlapping objects on two-dimensional  lattices.
Indeed, the constraint that any two particles appearing at the neighbouring sites have an infinite repulsion, (so that the input of such
configurations into the partition function of the adsorbate is zero), and the condition, in case of catalytically-activated reactions, that once two particles appear at the adjacent adsorption sites they react and leave the system, play essentially the same role.
A prominent example of such lattice gases,
solved exactly by Baxter \cite{baxter2,baxter}, is furnished by the so-called
hard hexagon model, which
is a two-dimensional lattice model of a gas, where identical particles are allowed to be on the
vertices of a triangular lattice but no two particles may be adjacent. From the perspective of catalytically-activated reactions, this model can be interpreted as a simple  reaction $A + A \to \oslash$ between monomers $A$, being at contact with a reservoir maintained at a constant chemical potential $\mu$, adsorbing onto empty lattice sites and desorbing back to the reservoir, and undergoing a chemical reaction as soon as any two $A$ particles appear at the adjacent sites.
Much progress has been made within the recent years
in understanding equilibrium properties of such lattice gases,
involving similar  or different particles, which may also have a different  shape (see, e.g., Refs. \cite{bouttier,brazil}), revealing a rich critical behaviour, characterised by continuous and discontinuous phase transitions which are quite analogous to the ones observed in the irreversible ZGB model.  We note also that a similarity between
a simple  $A + A \to \oslash$ reaction and random sequential adsorption has been pointed out in
Ref. \cite{satya}, since the removal of two nearest-neighbour reactants is equivalent to a "deposition" of a "dimer" of two empty sites. Such a duality has been used in Ref. \cite{satya} to derive an exact solution of the model on the Bethe lattice.

In this paper we study equilibrium properties of an adsorbate
formed in the course of
the reaction
\begin{equation}
A + A \to \oslash
\end{equation}
between monomers $A$, undergoing continuous exchanges with
their vapour phase - a reservoir maintained at a constant chemical potential $\mu$, and
taking place on the lattice with annealed
reactive properties.
Our goal here is to show that already this simplest
possible reaction scheme involving only monomers of the same type
exhibits quite a rich critical behaviour, depending moreover
on the specific way how
the reaction is modelled and also on the value of the particles' reactivity.
The critical behaviour which we predict is
characterised by a transition into a phase with
a broken symmetry (ordered or an  alternating state, in which layers with low and high densities are formed), which is continuous with a finite jump in compressibility. This transition is followed by, upon an increase of the chemical potential $\mu$
by a re-entrant transition into a disordered phase, which (depending on the reaction model)
may be
continuous with a finite jump in compressibility, or even discontinuous with a finite jump of density.

We will consider here two different ways of modelling the catalytic reaction:\\
In the Model I, we stipulate that
some fraction $p$ of the bonds $<ij>$ connecting nearest-neighbouring sites $i$ and $j$
possesses catalytic properties
so that the $A$ particles react when two of them appear on the neighbouring sites connected by such a bond.\\
 In the Model II,
we suppose that not the bonds but rather some lattice sites themselves are
the catalytic agents such that any two neighbouring $A$s will react instantaneously and leave the system, when at least one of them resides on a special catalytic sites.

More specifically, in the Model I we associate with each \textit{bond}  $<ij>$ a random variable
$\zeta_{<ij>}$, such that $\zeta_{<ij>} = 1$, if $<ij>$ is a catalytic bond
(which event is chosen with probability $p$, independently of other bonds) and
  $\zeta_{<ij>} = 0$, otherwise, (with probability $1-p$).
When any two $A$-s  appear
 at the neighbouring sites $i$ and $j$ connected by a catalytic bond,
 they  instantaneously react (and the product
 leaves the system). The $A$ particles harmlessly coexist on neighbouring
 sites connected by non-catalytic bonds.
For such a model, the corresponding grand  canonical partition function of the adsorbate with a fixed distribution of the catalytic bonds
 can be written as
\begin{equation}
\label{part1}
Z\left(\{\zeta_{<ij>} \}\right) = \sum_{\{n_j\}} \exp\left(\beta \mu \sum_j n_j\right) \, \prod_{<ij>} \Big(1 - \zeta_{<ij>} n_i n_j\Big) \,,
\end{equation}
where $\mu$  is the chemical potential, $\beta$ is the inverse temperature, (measured in units of the Boltzmann constant $k_B$), $n_j$ is the Boolean variable describing the occupation of a given site $j$ ($n_j = 1$ for an occupied site and $n_j=0$, otherwise),
the sum with the subscript $\{n_j\}$ denotes summation with
respect to the states of all occupation variables of all sites, and lastly,
the product sign with the subscript $<ij>$  signifies that
the product is taken over all bonds of the embedding lattice. For $p \equiv 1$, $Z\left(\{\zeta_{<ij>} \}\right)$ in eq. \ref{part1} evidentlycoincides
with the grand partition function of the gas of hard molecules with infinite nearest-neighbour repulsion \cite{baxter2,baxter}.

In the Model II, we assign to each of the \textit{sites} a random quenched variable $\chi_j$,
which, in a similar fashion, is equal
to $1$ for a "catalytic" site (probability $p$, independently of other sites)
and zero, otherwise, (with probability $1-p$). In this case, two $A$ particles
instantaneously react $A + A \to \oslash$, (and the product $\oslash$
 leaves the system), if at least one of them is on a catalytic site. If neither of them occupies a catalytic site, the particles do not react.
 For a given distribution of the catalytic sites, the grand canonical partition function
$Z\left(\{\chi_i \}\right)$ reads
\begin{equation}
\label{part2}
Z\left(\{\chi_i \}\right) = \sum_{\{n_j\}} \exp\left(\beta \mu \sum_j n_j\right) \, \prod_{i} \prod_{i, <ij>} \Big(1 - \chi_i n_i n_j\Big) \,,
\end{equation}
where the product with the subscript $i$ runs over all sites of the lattice,
while the product sign with the double subscript $i, <ij>$
signifies that the product operation is taken over
all the sites $j$ neighbouring to the site $i$. Note that the expression in eq. \ref{part2} becomes identical to the one in eq. \ref{part1} for $p \equiv 1$, as it should be.

We will focus in what follows  on the \textit{annealed} limit
in the distribution of the catalytic bonds or sites. Such a limit is interesting in its own right since it
corresponds to a physical situation in which the neighbouring adsorbed $A$ particles
may react at any place on the lattice, but only with a finite reaction probability $p$, so that the reaction between them is not perfect.
The properties of the adsorbate in this limit will be described by the annealed grand canonical partition functions, i.e., averaged
directly over the distributions of the random variables $\zeta_{<ij>}$ and $\chi_j$.
 The annealed grand canonical partition function for the Model I reads
\begin{eqnarray}
\label{part1av}
Z^{(\rm bonds)}(p) =  \sum_{\{n_j\}} \exp\left(\beta \mu \sum_j n_j\right) \, (1 - p)^{\sum_{<ij>} n_i n_j} \,,
\end{eqnarray}
and describes a grand canonical partition function of
a lattice gas with purely repulsive, \textit{soft} interactions.  Exact solution of the one-dimensional Model I for arbitrary $p$ for both the cases of
random quenched and of the annealed distribution of
the catalytic bonds has been obtained in Refs. \cite{bur1,bur2}.

In turn, for the annealed Model II we have  the following grand canonical
partition function
\begin{equation}
\label{deff}
Z^{(\rm sites)}(p) = \sum_{\{n_j\}} \exp\left(\beta \mu \sum_i n_i\right) \,
(1-p)^{\sum_i n_i \Psi_i } \,,
\end{equation}
where $\Psi_i$ is a Boolean function of the form
\begin{equation}
\label{bool}
\Psi_i = 1 -  \prod_{i, <ij>} \left(1 - n_j\right) \equiv
\cases{0 &when all $n_j =  0$,\\
1&when at least one $n_j = 1$.\\}
\end{equation}
Therefore, in the annealed Model II we again deal with a lattice gas with purely repulsive, \textit{soft} interactions between the particles, but
here the interactions are very peculiar: the total interaction energy of the particle occupying site $i$
is not proportional
to the number of neighbouring particles, as it usually happens,
but has a Boolean form such that $\Psi_i = 0$
if all the sites neighbouring to $i$ are vacant, and $\Psi_i = 1$ independently of the actual number of occupied neighbouring sites
 if at least one of them  is
 occupied.
 Note that in this rather exotic model of statistical mechanics the interactions are therefore less restrictive than those appearing in the Model I, in which each pair of neighbouring particles contributes to the energy of the system.  Exact solution of the Model II in one-dimensional systems was
obtained in Refs. \cite{ol1,ol2} for arbitrary $p$,
for both the cases of random quenched and of the annealed distributions of the catalytic sites.

Analysing the critical behaviour of the \textit{annealed} versions
 of the Model I and Model II,
we will resort here to a mean-field-like approximation, solving both models
exactly on the Bethe lattice  - a deep interior of the so-called Cayley tree (see Fig.\ref{Fig1}, left panel) well away of the boundary sites. This is a topologically simpler object
than usual lattices, because such trees do not have closed loops (such that the approximation is equivalent to the standard Bethe-Peierls theory \cite{baxter}).
We focus solely on the case of a Bethe lattice with the coordination number equal to three, which means that we provide here an approximate solution for the annealed Models I and II on a honeycomb lattice.
We note that
 for many complicated lattice
models of statistical mechanics such an approximation has been invoked first, since it is often amenable to a completely analytical analysis reducing the original many-body
problem to the analysis of the limit solutions of some non-linear recursion schemes,
in which a phase transition manifests itself by the spontaneous break up of the symmetry between
recursion terms of an odd and an even order. In many cases,
a detailed studies of the properties of the original lattice models can be carried out,
showing  no sign of a pathological behaviour \cite{gujrati}. To name but a few, we mention
studies of the phase diagrams of athermal lattice gases \cite{runnels,muller,brazil},
modulated phases of the Ising model
with competing interactions \cite{van,hor,mariz}, Potts models \cite{potts}, lattice models of glassy systems \cite{biroli1,biroli2}, different aspects of the localisation transition \cite{biroli3,biroli5} and phase diagram
for ionic liquids in non-polarised nano-confinement \cite{kondrat}.

Our \textit{main} results are as follows:\\
For the annealed Model I defined on the Bethe lattice with the coordination number three, we set out to show that there exists\\ a) a critical value $p = p_c^{(bonds)} = 8/9$, such that for $p < p_c^{(bonds)}$ no broken symmetry phase emerges, while for $p_c^{(bonds)} < p < 1$, upon increasing the chemical potential $\mu$, one observes\\ b) a transition from a disordered to an ordered (alternating) phase at $\mu = \mu_{c,1}^{(bonds)}(p)$. In this ordered phase the tree spontaneously partitions into alternating layers - the ones which are densely populated by the $A$ particles and the ones which are almost completely devoid of them.
For the original honeycomb lattice it will mean that (somewhere close to $\mu = \mu_{c,1}^{(bonds)}(p)$) the lattice will partition into two sub-lattices - the one densely packed with $A$ particles and the second - almost empty.\\ c) Increasing $\mu$ further,  a re-entrant (inverted) transition into a disordered phase at $\mu = \mu_{c,2}^{(bonds)}(p) > \mu_{c,1}^{(bonds)}(p)$ should take place. Both transitions are continuous, with a finite jump in compressibility.  When $p \to 1$, the re-entrant transition disappears since $\mu_{c,2}^{(bonds)}(p) \to \infty$.\\
d) We also calculate the mean density in the disordered and ordered phases,
the staggered density in the ordered phase which serves as the order parameter for the system under study and the compressibility.\\
For the annealed version of the Model II defined on the Bethe lattice with the coordination number three, we observe a critical behaviour, which turns out to be even somewhat richer than that of the Model I.
We show that\\ a) there exist two critical values of the reaction probability $p$: $p_{c,1}^{(sites)} \approx 0.794$ and  $p_{c,2}^{(sites)} \approx 0.813$ and no critical behaviour emerges for $p < p_{c,1}^{(sites)}$.\\ b) For $p > p_{c,1}^{(sites)}$, upon a gradual increase of $\mu$, the system undergoes a continuous transition from a disordered into an ordered, alternating phase at $\mu = \mu_{c,1}^{(sites)}(p)$, characterised by a finite jump of the compressibility.\\ c) Our further analysis reveals that the behaviour for larger values of $\mu$ depends on whether $p$ belongs to the interval $p_{c,1}^{(sites)} < p < p_{c,2}^{(sites)}$ or to the interval $p \geq p_{c,2}^{(sites)}$: for the former case, for $\mu = \mu_{c,2}^{(sites)}(p) > \mu_{c,1}^{(sites)}(p)$ the system undergoes a re-entrant \textit{continuous} transition into a disordered phase, while for the latter  this transition is \textit{discontinuous}, with a finite jump of both the mean and the staggered densities.\\ d)  We also determine the mean density in the disordered and the ordered phases, the staggered density and the compressibility.

The paper is outlined as follows: In Sec. \ref{model} we describe the system's geometry
and present the details of the
derivation of the annealed grand canonical partition functions (in what follows, we will call them for brevity just as "partition functions") for the Models I and II. In Sec. \ref{Runnels} we focus on the singular case $p \equiv 1$, in which both Models  (as well as their annealed versions) are identical, and briefly recall the classical analysis
due to Runnels of the critical behaviour of a gas of hard molecules on the Bethe lattice with the coordination number equal to three.
This permits us to set up a unifying framework for what follows.
In Sec. \ref{m1} we study the annealed version of the Model I for an arbitrary value of the reaction probability $p$, $0 < p \leq 1$. We first derive the recursion relations obeyed by the partition function in eq. \ref{part1av} on a Bethe lattice  and analyse their
critical behaviour, as manifested by the breaking of the symmetry between the terms of odd and even order.
Next, in Sec. \ref{m2}, we focus on the annealed version of the Model II. We derive here the recursion relations obeyed by the annealed  partition function defined in eqs. \ref{deff} and \ref{bool}, and analyse their critical behaviour, and also determine the mean density in the disordered and ordered phases, the staggered density and the compressibility, as the functions of the chemical potential. Finally, we conclude in Sec. \ref{conc} with a brief recapitulation of our results.

\section{\label{model} Partition functions in the annealed limit}

\begin{figure}
\includegraphics[width=.45\hsize]{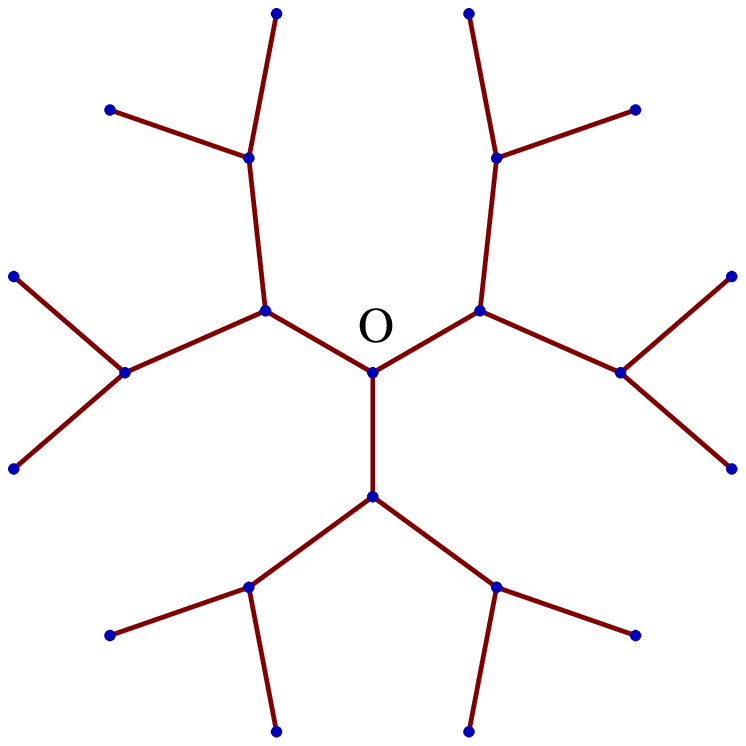}
\includegraphics[width=.45\hsize]{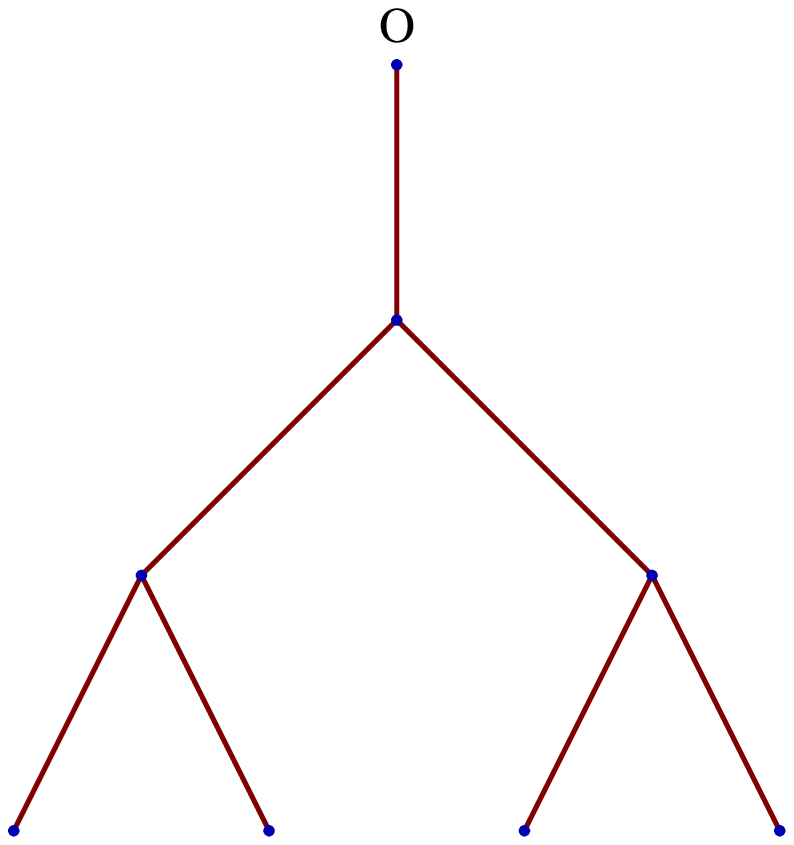}
\caption{ Left: Cayley tree with $3$ generations emanating from the central site ${\rm O}$. Right: A subtree with $3$ generations emanating from the central site ${\rm O}$. }
\label{Fig1}
\end{figure}

Consider  a Cayley tree with the coordination number equal to three,  (see Fig.\ref{Fig1}, left panel), having
$N$ generations  of sites (such that the "volume" $M = 3 (2^N - 1) + 1$ for $N \geq 0$ ), in contact with the vapour phase of hard
particles $A$.
The $A$ particles can
adsorb onto the vacant sites and can desorb back to the reservoir.
The reservoir is characterised
by a chemical potential  maintained at a constant value $\mu$
and measured relative to the binding energy of an occupied site, such that $\mu > 0$
indicates a preference for adsorption. Correspondingly,
the activity $z$ is defined as $z = \exp(\beta \mu)$.
Hard-core interactions prohibit
 double occupancy of any node $j$ and the occupation of this node is described by a Boolean variable $n_j$ such that
 $n_j = 1$ for an occupied node and $n_j=0$,
 otherwise.

We will consider  two ways of modelling the catalytic reaction between the neighbouring $A$ particles,
which differ in the definition of the catalytic agent. In Model I, we suppose that
these are some bonds connecting neighbouring sites, which are deemed to have
special catalytic properties promoting the reaction, while in Model II these will be some special catalytic sites. In the former case, only the $A$ particles separated by a catalytic bond react, while in the latter, in order for the reaction to take place, at least one of two neighbouring $A$s has to reside on a special catalytic site. In both cases, a given bond or a given site can be in a catalytic state with probability $p$, independently of the environment.
The  partition functions of the adsorbate  corresponding to the Models I and II, $Z\left(\{\zeta_{<ij>} \}\right)$ and $Z\left(\{\chi_i \}\right)$, are defined in eqs. \ref{part1} and \ref{part2}, respectively.

Here we will be concerned with the annealed versions of both Models whose partition functions are obtained by directly averaging $Z\left(\{\zeta_{<ij>} \}\right)$ and $Z\left(\{\chi_i \}\right)$ over the distributions of the catalytic bonds and sites.

The averaging of $Z\left(\{\zeta_{<ij>} \}\right)$ in eq. \ref{part1} is very straightforward
\begin{eqnarray}
\label{part1avav}
 Z^{(\rm bonds)}(p) &=& \left \langle Z\left(\{\zeta_{<ij>} \}\right) \right \rangle_{\zeta_{<ij>} } = \sum_{\{n_j\}} z^{\sum_j n_j} \, \prod_{<ij>} \Big\langle \Big(1 - \zeta_{<ij>} n_i n_j\Big) \Big\rangle_{\zeta_{<ij>} } \nonumber\\
\fl &=& \sum_{\{n_j\}} z^{\sum_j n_j} \, \prod_{<ij>} \Big( p \left(1 - n_i n_j\right) + 1 - p\Big) \,,\nonumber\\
\fl &=&  \sum_{\{n_j\}} z^{\sum_j n_j} \, (1 - p)^{\sum_{<ij>} n_i n_j} \,,
\end{eqnarray}
from which one can directly read off  the result in eq. \ref{part1av}. Averaging of $Z\left(\{\chi_i \}\right)$ in eq. \ref{part2} is only slightly more involved. Here we have
\begin{eqnarray}
 Z^{(\rm sites)}(p) &=& \langle Z\left(\{\chi_i \}\right)  \rangle_{\chi_i} = \sum_{\{n_j\}} z^{\sum_j n_j} \, \prod_{i}  \, \Big\langle \prod_{i, <ij>} \Big(1 - \chi_i n_i n_j\Big)\Big\rangle_{\chi_i} \nonumber\\
\fl &=& \sum_{\{n_j\}} z^{\sum_j n_j} \, \prod_{i}  \, \left( p \prod_{i, <ij>}  \left(1 - n_i n_j\right) + 1 - p\right) \,.
\end{eqnarray}
Further on, noticing that
\begin{equation}
 \left( p \prod_{i, <ij>} \left(1 - n_i n_j\right) + 1 - p\right) \equiv  \left( p \prod_{i, <ij>} \left(1 - n_j\right) + 1 - p\right)^{n_i} \,,
\end{equation}
and next
that, trivially,
\begin{equation}
 \left( p \prod_{i, <ij>} \left(1 - n_j\right) + 1 - p\right) \equiv (1-p)^{1 - \prod_{i, <ij>} \left(1 - n_j\right)} \,,
\end{equation}
we find that the disorder-average partition function $Z^{(\rm sites)}(p)$ attains the form defined in our eqs. \ref{deff} and \ref{bool}.
Parenthetically, we note that the expressions in eqs. \ref{part1av} and \ref{deff} are valid
for any lattice, not necessarily for the Cayley tree.

Next, focussing on the limit $N \to \infty$, we will discard
the surface effects considering the behaviour in the interior of the Cayley tree. In other words, we will turn
to the thermodynamic limit of the Cayley tree, i.e., to the behaviour on the so-called Bethe lattice.
Under such simplifying assumptions,
 the thermodynamics of the system is given by the annealed
 pressure (in units of the lattice cell area), defined as
\begin{equation}\label{pressure0}
P_I(T,\mu) = \frac{1}{\beta} \lim_{M \to \infty} \frac{\ln Z^{(bonds)}(p)}{M} \,,
\end{equation}
for the Model I, and as
\begin{equation}
\label{pressure2}
P_{II}(T,\mu) = \frac{1}{\beta} \lim_{M \to \infty} \frac{\ln Z^{(sites)}(p)}{M} \,,
\end{equation}
for the Model II.  Once $P$-s are known, all other thermodynamic quantities of interest can be
obtained by differentiating $P$ with respect to $\mu$ or $T$.

\section{\label{Runnels} Singular case $p \equiv 1$. Runnels' analysis.}

To set up the scene, we first consider the special singular case $p \equiv 1$, in which all the bonds and all the sites are catalytic, implying  that  the reactive constraint is imposed everywhere. Analysis in this case is much more simple than for an arbitrary $p$ but will permit us to introduce all the basic concepts
and to establish a unifying framework for what follows.
In this case, eqs. \ref{part1} and \ref{part2} become identical, and also identical to the partition function of
the lattice gas of hard particles, having an infinite repulsion between the neighbouring ones:
\begin{equation}
Z(\{1\}) = \sum_{\{n_j\}} z^{\sum_j n_j} \, \prod_{<ij>} \Big(1 - n_i n_j\Big) \,.
\end{equation}
Below we briefly recall some classical results due to Runnels \cite{runnels} obtained for $Z(\{1\})$ defined on the Bethe lattice with the coordination number three, using for further convenience
a bit different settings and notations.

The  partition function  $Z(\{1\})$ of the entire Cayley
tree writes:
\begin{equation}
\label{2}
Z(\{1\}) =  Z_N^{(0)}(\{1\}) + Z_N^{(1)}(\{1\})\,,
\end{equation}
where $Z_N^{(0)}(\{1\})$ and $Z_N^{(1)}(\{1\}$ are the partition functions of the entire tree with a vacant and an occupied central site, respectively.

Define next a
subtree with $N$ generations (see Fig. \ref{Fig1}, right panel), emanating from the central site $O$,
and introduce two auxiliary partition functions -
 $B_N(1)$ and $B_N(0)$,
 where the former denotes the partition function of a subtree
 with an occupied central site, while for
 the latter the central site of the subtree
 is vacant.
Clearly, we have
\begin{equation}
\label{m}
Z_N^{(0)}(\{1\}) = B_N^3(0) \,,
\end{equation}
because the entire Cayley
 tree with the vacant central site decomposes into three independent subtrees with a vacant central site,
while  $Z_N^{(1)}(\{1\})$ decomposes into three independent subtrees with an occupied central site :
\begin{equation}
\label{mm}
Z_N^{(1)}(\{1\}) = z^{-2} B_N^3(1) \,,
\end{equation}
where the factor $z^{-2}$ prevents the over-counting of the contribution of the central site. We seek next the recursion relations obeyed by
$B_N(1)$ and $B_N(0)$. For $p \equiv 1$, if the central site is occupied, the neighbouring one (generation $1$)
is always vacant, such that
\begin{equation}
\label{9}
B_N(1) = z \, B_{N - 1}^2(0) \,,
\end{equation}
where the factor $z$ stems from
 the contribution of the occupied central site. Next, if the central site is vacant, the neighbouring one (generation $1$)
can be either vacant or occupied, such that
$B_N(0)$ can be represented via the  partition functions of the subtrees with $N-1$ generations as
\begin{equation}
\label{10}
B_N(0) = B_{N - 1}^2(0) + z^{-1} \, B_{N - 1}^2(1) \,,
\end{equation}
where the factor $z^{-1}$ in the second term on the r-h-s of eq. \ref{10} prevents the over-counting of the contribution of the occupied site,
neighbouring the central one, from which
 two subtrees with $N-1$ generations
emanate. Together with the evident "initial" conditions $B_0(1) = z$ and $B_0(0) = 1$, eqs. \ref{9} and \ref{10} totally define $B_N(1)$ and $B_N(0)$ (and hence, $Z(\{1\})$), for the Cayley tree with an arbitrary number of generations $N$.

Equations \ref{9} and \ref{10} can be simplified by introducing an auxiliary parameter
\begin{equation}
\label{11}
x_N(z) = \frac{B_N(1)}{z B_N(0)} \,.
\end{equation}
This parameter, multiplied by $z$, is simply
the ratio of the number of the
subtrees with an occupied central site
and the number of the subtrees with a vacant central site.  Dividing eq. \ref{9} by eq. \ref{10}, we find that $x_N(z)$ obeys the following recursion :
\begin{equation}
\label{rec1}
x_N(z) = \frac{1}{1 + z \, x_{N-1}^2(z)}\,, \,\,\, x_0(z) = 1 \,.
\end{equation}
Once $x_N(z)$-s are determined,
one can readily find $B_N(0)$ via the recursion
\begin{equation}
\label{k}
B_N(0) = \left(1 + z x^2_{N-1}(z)\right) B_{N-1}^2(0)= \frac{B_{N-1}^2(0)}{x_{N}(z)}\,, \,\,\, B_0(0) = 1 \,,
\end{equation}
which follows from eq. \ref{10} and the definitions of $x_N(z)$ in eqs. \ref{11} and \ref{rec1}, and gives
\begin{equation}
B_N(0) = \prod_{j = 0}^{N} \frac{1}{x^{2^{N  - j}}_{j}(z)} \,.
\end{equation}
In turn,
the partition function of the entire Cayley tree in eq. \ref{2} can be straightforwardly written using $x_N(z)$-s as
\begin{equation}
Z_N(\{1\}) =  \left(1 + z x_N^3(z)\right) \left(\prod_{j = 0}^{N} \frac{1}{x^{2^{N  - j}}_{j}(z)}\right)^3 \,,
\end{equation}
so that the pressure $P$ in the limit $N \to \infty$ (and hence, $M \to \infty$) will be given by
\begin{eqnarray}
\label{pressure}
P(T,\mu) &=& \frac{1}{\beta} \lim_{N \to \infty} \left(3 \frac{2^N}{M} \sum_{j=0}^{N } 2^{- j } \ln\left(\frac{1}{x_{j}(z)}\right) + \frac{\ln\left(1 + z x_N^3(z)\right)}{M}  \right) \nonumber\\
\fl &=&
\frac{1}{\beta} \sum_{j=0}^{\infty} 2^{- j } \ln\left(\frac{1}{x_{j}(z)}\right) \,.
\end{eqnarray}
Note that eq. \ref{pressure} defines the pressure of the adsorbate on the entire Cayley tree, such that to get an analogous result for the Bethe lattice we have to subtract the contribution due to the boundary sites.
For the case $p\equiv 1$, the corresponding procedure will be explained below. For the general case $0 < p \leq 1$  we will adapt the procedure elaborated in Ref. \cite{ananikian}, which is explained in \ref{A2}.

\begin{figure}
\begin{center}
\includegraphics[width=.6\hsize]{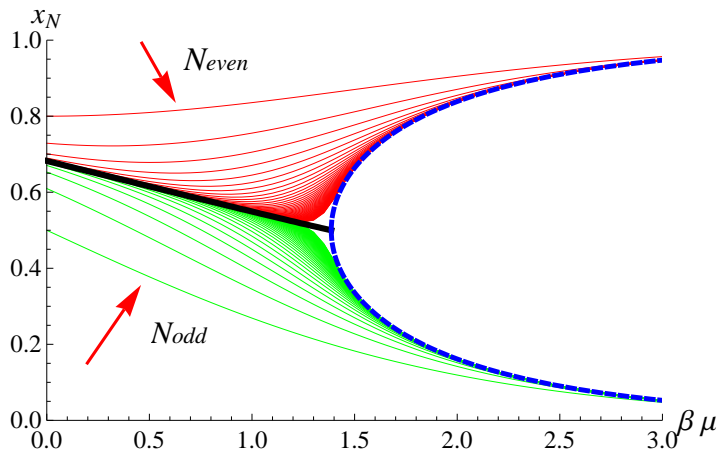}
\caption{ $x_N=x_N(z)$ in eq. \ref{rec1} for $N=1,2,\ldots, 100$
plotted versus $\beta \mu = \ln(z)$. Red curves are $x_N(z)$-s with even $N$, while  the green ones correspond to the terms with odd values of $N$. Arrows indicate the direction of growth of $N$.
Thick black curve depicts
the behaviour of the real root of
eq. \ref{small} (see eq. \ref{real}) on the interval $[0,\beta \mu = 2 \ln(2)]$. Two dashed blue curves starting at $\beta \mu = 2 \ln(2)$ are the roots of the second term in eq. \ref{roots} defining two distinct limit curves $x_{\rm odd}(z)$ and $x_{\rm even}(z)$. The figure illustrates continuous phase transition from disordered state described by solution of eq. \ref{small} (black line) to symmetry broken phase with alternating layers of different densities described by solution of eq. \ref{roots} (dashed blue lines). The phase transition occurs at $z_c=4$ at which all three curves meet.
}
\end{center}
\label{Fig3}
\end{figure}

\subsection{Solutions of the recursion relations for $p \equiv 1$}

The central question now is the behaviour of $x_N(z)$-s, which are all functions of the activity $z$.
Before we turn to analytical calculations, it might be expedient to numerically  generate several consecutive
terms $x_N(z)$, in order to get a clue to their actual behaviour.
 In Fig. 2 we present first $100$  terms $x_N(z)$ as functions of the activity $z = \exp(\beta \mu)$. One observes
 that for a relatively small $\beta \mu$ (or $z$), as $N$ grows,
all $x_N(z)$ converge to the same curve $x(z)$. Therefore,
for sufficiently small $z$ one has a sequence convergence as $N \to \infty$
to the unique limit $x(z)$, which, in virtue of eq. \ref{rec1}, obeys
\begin{equation}
\label{small}
z \, x^3(z) + x(z) - 1 = 0 \,.
\end{equation}
The discriminant $\triangle = - 27 z^2 - 4 z$ of the cubic eq. \ref{small}
is strictly negative, meaning that eq. \ref{small} has two non-real complex conjugate roots and one real root $x(z)$ which
is given explicitly by
\begin{equation}
\label{real}
x(z) = - \frac{1}{3 z} \left(C + \frac{\triangle_0}{C}\right)\,,
\end{equation}
where
\begin{equation}
\triangle_0 = - 3 z \,,    C =  \left(\frac{\triangle_1 + 3 z \sqrt{- 3 \triangle}}{2}\right)^{1/3} \,, \triangle_1 = - 27 z^2  \,.
\end{equation}
This root is depicted in Fig.2 by a thick black curve.

Further on, Fig. 2 shows that there is an apparent
"critical" value $z_c$ such that for $z > z_c$ the odd terms converge to one function, $x_{2 N + 1}(z) \to x_{\rm odd}(z)$, while even terms converge to another one, $x_{2 N}(z) \to x_{\rm even}(z)$. In this case of the so-called subsequence convergence, one may calculate $x_{\rm odd}(z)$ and $x_{\rm even}(z)$  by iterating eq. \ref{rec1} once again, so that the the resulting recursion scheme will involve the terms of the same parity only; that is, we rewrite eq. \ref{rec1} formally as
\begin{equation}
\label{rec2}
x_N(z) = \left(1 + \frac{z}{\left(1 + z \, x^2_{N - 2}(z)\right)^2}\right)^{-1}\,.
\end{equation}
Turning next to the limit $N \to \infty$, we find that two distinct limit functions $x_{\rm odd}(z)$ and $x_{\rm even}(z)$ must satisfy the fifth-order equation, that conveniently factors into
\begin{equation}
\label{roots}
\Big(z \, x^3(z) + x(z) - 1\Big) \Big(z x^2(z) - z x(z) + 1\Big) = 0 \,.
\end{equation}
One notices that the first term is precisely our previous eq. \ref{small}, which has a real root defined by eq. \ref{real}, while the second term has two roots:
\begin{equation}
\label{even}
x_{\rm even}(z) = \frac{z + \sqrt{z^2 - 4 z}}{2 z} \,,
\end{equation}
and
\begin{equation}
\label{odd}
x_{\rm odd}(z) = \frac{z - \sqrt{z^2 - 4 z}}{2 z} \,,
\end{equation}
which we depict in Fig. 2 by blue dashed curves.
These two roots are real only for $z \geq 4$. Note now that the three curves defined by eqs. \ref{real}, \ref{even} and \ref{odd}, all meet at $z = 4$ each assuming the value $1/2$. Since $x_{\rm odd}(z)$ and $x_{\rm even}(z)$ are complex-valued for $z < 4$, the unique limit $x(z)$ in eq. \ref{real} must obtain for $z < 4$. As shown in Ref. \cite{runnels}, the result in eq. \ref{real} becomes unstable for $z > 4$. This signifies that the alternating limits  $x_{\rm odd}(z)$ and $x_{\rm even}(z)$ obtain for $z > 4$.

One may notice that 
the special point $z=4$, 
where the singularity occurs, can be obtained from the direct stability analysis 
of the fixed point solution of recursion \ref{rec1}. This recursion is stable if the slope of the 
function $f(x_{N-1}(z))$ in the right-hand-side of eq. \ref{rec1} in the vicinity of the 
fixed point is smaller than unity; that being $|f'(x_{N-1}(z))|_{x_{N-1}(z)=x(z)}<1$. Therefore,  solving equation:
\begin{equation}
\left.f'(x_{N-1}(z))\right|_{x_{N-1}(z)=x(z)}=-1
\end{equation}
with $x(z)$ defined by eq. \ref{real} one gets $z=4$ as the point where the solution \ref{real} looses its stability and a bifurcation occurs. For $z>4$ the limit cycle defined by the second term in eq. \ref{roots} is a stable solution.

Therefore, $z = 4$ is the only candidate for a singularity or a phase transition. As shown by Runnels \cite{runnels} (see also Ref. \cite{muller} for a more detailed discussion of the behaviour on the entire Cayley tree), in actual fact it is a rather subtle issue. The point is that on the Cayley tree the various derivatives of $x_j$ become indeed unbounded with increasing $j$,
but not rapidly enough to spoil the convergence of eq. \ref{pressure} and its derivatives, due to the presence of the factor $2^{-j}$. The ensuing smooth overall behaviour attributable to this factor is easily understood : most of the sites of the Cayley tree
are near its exterior surface, where  they are essentially independent; the highly correlated sites are deep in the interior of the tree. In other words, zeros of the  partition function do close in on $z = 4$ but with vanishing density, which allows the system to sneak through on the real axis with no transition. Conversely,  if one discards the influence of the overwhelming majority of the
surface sites and
focuses on the behaviour of the interior of the Cayley tree far away the surface - the so-called Bethe lattice, one will indeed observe a transition at $z = z_c = 4$.
The transition occurring in this interior region far removed from the exterior
surface is continuous, with a finite jump in the compressibility \cite{runnels}.
This is a transition of the disorder-order type so that at $z = z_c = 4$ the interior region spontaneously partitions in alternating, highly occupied and almost devoid of particles
layers.  Note that these predictions on the existence and on the type of the phase transition
obtained for the Bethe lattice, which is essentially a mean-field (albeit quite reliable \cite{gujrati}) approximation,
appear to be qualitatively correct for models defined on corresponding (with the same coordination number) regular lattices, as evidenced later by
the exact solution of the so-called hard-hexagons model  \cite{baxter2}.

\subsection{Order parameter for $p \equiv1$}

We briefly recall next Runnel's calculations  of the particle densities in the alternating layers. Differentiating eq. \ref{pressure}, one gets for the mean density
 \begin{equation}
 \label{density}
 \rho(z) = \beta z \frac{\partial P}{\partial z}  = \sum_{j = 1}^{\infty} 2^{-j} t_j(z) \,,
 \end{equation}
 where $t_j(z) = - z \partial \ln\left(x_j(z)\right)/\partial z$ satisfies the recursion of the form
 \begin{equation}
 \label{tj}
 t_j(z) = b_j(z) \left(1 - 2 t_{j-1}(z)\right) \,, \,\,\, t_0(z) = 1 \,,
 \end{equation}
 in terms of $b_j(z) = 1 - x_j(z)$. Straightforward calculations give \cite{runnels}
 \begin{equation}
 \label{comp}
 t_j(z) = \sum_{i = 0}^{j-1} (-2)^j \prod_{n = j - i}^{j} b_n(z) \,,
 \end{equation}
 which, being inserted in eq. \ref{density}, allows to rewrite the latter equation as
 \begin{equation}
 \label{exp}
 \rho(z) = \sum_{j = 0}^{\infty} 2^{-j} r_j(z) \,,
 \,\,\,
 r_j(z) = \sum_{i=0}^{\infty} (-1)^{i} \prod_{n=j}^{j+i} b_n(z) \,.
 \end{equation}
Note now that $b_j(z)$ and $r_j(z)$ have simple interpretations \cite{runnels}: From the preceding equations, it is clear that $b_j(z)$ is the fraction of the activity-weighted configurations of subtrees having an occupied root, or in other words, the average occupancy of the root at activity $z$. Noticing then that
 eq. \ref{exp} can be alternatively written as
 \begin{equation}
 \label{occ}
 r_j(z) = b_j(z) \left(1 - r_{j+1}(z)\right) \,,
 \end{equation}
 (which can be straightforwardly reiterated to give back eq. \ref{exp}), one concludes that $r_j(z)$ is simply the average occupancy of a site at generation $j$. Indeed,  eq.  \ref{occ} can be interpreted as the definition of
 the average occupancy $r_j(z)$: it equals the factor $b_j(z)$, that being, the weighted fraction of configurations of a subtree with the selected site occupied, times the factor $(1- r_{j+1}(z))$, which defines the fraction of configurations of a subtree with the inner site adjacent to the selected site vacant to permit occupancy of the selected site. Therefore, eq.  \ref{exp} shows that the overall density is a weighted average of the densities of various generations, the weighting factor $2^{-j}$ being proportional to the number of sites in generation $j$.

One may now define the order parameter as the
staggered density $\delta \rho = |r_j(z) - r_{j+1}(z)|$, i.e., the difference of average occupations of adjacent layers deep in the interior of the tree
 in the limit $j \to \infty$.  Clearly, for $z < z_c$ all $b_j(z)$ converge to the single limit $b(z) = 1 - x(z) $ regardless of the parity, where $x(z)$ is the solution of eq. \ref{small}. In this case, in virtue of eq. \ref{occ}, the average occupations of the layers converge as $j \to \infty$
 to the unique limit $r_j(z) \to r(z) = b(z)/(1 + b(z))$ such that $\delta \rho \equiv 0$.

For $z > z_c$ the situation is different. As $j \to \infty$, the parameter $b_j$ with odd $j$ converges to $b_{+}(z) = 1 - x_{odd}(z)$, eq. \ref{odd},
while the parameter $b_j$ with even $j$ converges to $b_{-}(z) = 1 - x_{even}(z)$, eq. \ref{even}. Consequently, one finds from eq. \ref{exp} that the densities $r_j(z)$ in the layers with odd $j$ converge to $r_+(z)$, while the average occupations of the layers with even $j$ converge to $r_{-}(z)$, which are given explicitly by
\begin{equation}
\label{+}
r_+(z) = \frac{b_+(z) (1 - b_-(z))}{1 - b_+(z) b_-(z)} \,, r_-(z) = \frac{b_-(z) (1 - b_+(z))}{1 - b_+(z) b_-(z)} \,.
\end{equation}
In turn, for $z > z_c = 4$ the order parameter obeys
\begin{equation}
\delta \rho = \frac{b_+(z) - b_-(z)}{1 - b_+(z) b_-(z)} = \frac{x_{even}(z) - x_{odd}(z)}{x_{even}(z) + x_{odd}(z) - x_{even}(z) x_{odd}(z)} \,.
\end{equation}
and equals zero for $z < z_c$. In the limit $z \to \infty$, $r_+(z) \to 1$, $r_-(z) \to 0$, and $\delta \rho \to 1$.

\section{Model I : Catalytic bonds}
\label{m1}

In this Section we analyse the critical behaviour of the annealed version of the Model I, whose   partition function is defined by eq. \ref{part1av}.
Here, similarly to the previously considered case $p \equiv1$, $Z^{(\rm bonds)}(p)$ can be formally decomposed as
$Z^{(\rm bonds)}(p) = Z_N^{(\rm bonds,0)}(p) + Z_N^{(\rm bonds,1)}(p) $, where
now   $Z_N^{(\rm bonds,0)}(p)$ ($ Z_N^{(\rm bonds,1)}(p)$) denotes the
disorder-average  partition functions of the Cayley tree with a vacant (occupied) central site.
In this case eqs. \ref{m} and \ref{mm} read
\begin{equation}
\label{b0}
Z_N^{(\rm bonds,0)}(p) = B_N^3(0,p) \,, Z_N^{(\rm bonds,1)}(p) =z^{-2} B_N^3(1,p) \,,
\end{equation}
where $B_N(0,p)$ and $B_N(1,p)$ are the partition functions of a subtree (see, Fig. \ref{Fig1}, right panel) with a vacant and an occupied root, respectively.   Further on, we find straightforwardly that $B_N(0,p)$ and $B_N(1,p)$ obey the following recursions :
\begin{equation}
\label{33}
B_N(1,p) = z B_{N-1}^2(0,p) + z (1-p) \frac{B_{N-1}^2(1,p) }{z} \,,
\end{equation}
and
\begin{equation}
\label{333}
B_N(0,p) = B_{N-1}^2(0,p) + z^{-1} B_{N-1}^2(1,p) \,.
\end{equation}
While the latter expression coincides with eq. \ref{10} of the previous section, the former one, given by eq. \ref{33}, has a different form compared to that in eq. \ref{9} since here it is possible to have two particles at the neighbouring sites.

The recursions in eqs. \ref{33} and \ref{333} can be simplified by introducing
\begin{equation}
\label{xx}
x_N(p,z) = \frac{B_N(1,p)}{z B_N(0,p)} \,,
\end{equation}
which now obeys the recursion :
\begin{equation}
\label{rec3}
x_N(p,z) = \frac{1 + z (1 - p) x_{N-1}^2(p,z)}{1 + z  x_{N-1}^2(p,z)} \,,\,\,\, x_0(p,z) = 1\,.
\end{equation}
Note that $B_N(0,p)$ still obeys $B_N(0,p) = (1 + z x_{N-1}^2(p,z)) B^2_{N-1}(0,p)$, like the analogous property in the case $p \equiv1$, eq. \ref{k}, which implies that the pressure $P_I(T,\mu) $ of the adsorbate is given by last line of eq. \ref{pressure} with $x_j(z)$ replaced by $(x_j(p,z)-1+p)/p$.

\subsection{Solutions of the recursion relations for $p < 1$}

We proceed further by numerically iterating first $20$ terms of the recursion defined in eq. \ref{rec3} for two different values of the parameter $p$: $p = 0.85$ and $p = 0.95$,  which we present in panels (a) and (b) of Fig. 3. For $p = 0.85$ (panel (a))
we observe a sequence convergence $x_N(p,z) \to x(p,z)$, as $N \to \infty$,
to a single limiting curve $x(p,z)$ for both even and odd $N$ for any $z$. For this single limit case, $x(p,z)$ obeys,  in virtue of eq. \ref{rec3},
a cubic equation of the form,
\begin{equation}
\label{p1}
z x^3(p,z) - z (1- p) x^2(p,z) + x(p,z) - 1 = 0 \,,
\end{equation}
which reduces to eq. \ref{small} when $p = 1$. The discriminant $\triangle(p)$ of the cubic eq. \ref{p1},
\begin{equation}
\triangle(p) = - 4 (1 - p)^3 z^3 - (8 + 20 p - p^2) z^2 - 4 z \,,
\end{equation}
is strictly negative, so that eq. \ref{p1} has a single real root given explicitly by
\begin{equation}
\label{singleroot}
x(p,z) = \frac{1 - p}{3} - \frac{1}{3 z} \left(C(p) + \frac{\triangle_0(p)}{C(p)}\right) \,,
\end{equation}
where
\begin{equation}
\triangle_0(p) = (1 - p)^2 z^2 - 3 z \,, C(p) = \left(\frac{\triangle_1(p) + 3 z \sqrt{-3 \triangle(p)}}{2}\right)^{1/3} \,,
\end{equation}
and
\begin{equation}
\triangle_1(p) = - z^2 \left(9 (2 + p) + 2 (1 - p)^3 z\right) \,.
\end{equation}
The real root in eq. \ref{singleroot} is depicted by a thick black line in Fig. 3, panel (a).

\begin{figure}
\begin{center}
\includegraphics[width=.48\hsize]{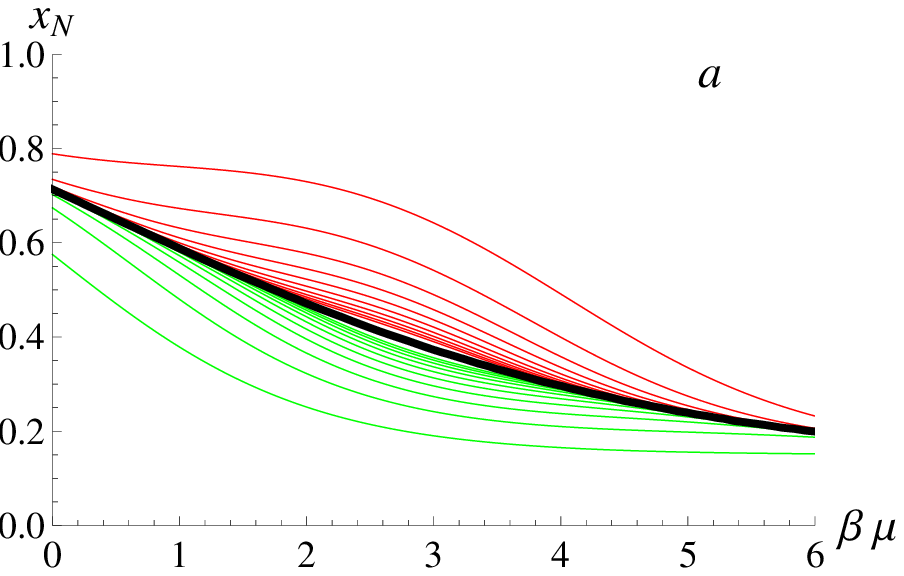}
\includegraphics[width=.48\hsize]{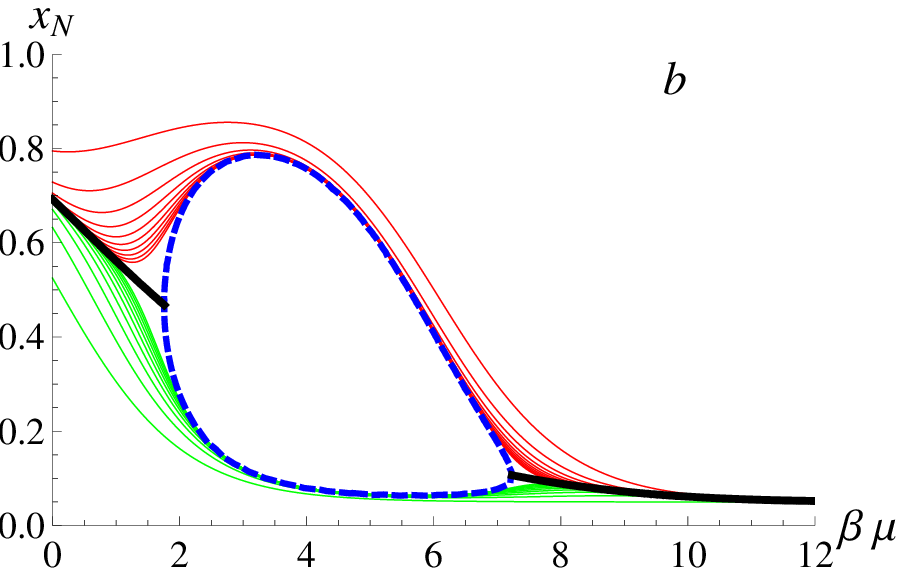}
\caption{ $x_N=x_N(p,z)$ in eq. \ref{rec3} for $N=1, 2, \ldots, 20$
plotted versus $\beta \mu = \ln(z)$. Red curves are $x_N$ with even $N$, while  the green ones correspond to odd values of $N$. Panel (a): $p = 0.85$.
Thick black curve describe disordered state and is
 the real root of
eq. \ref{p1} given explicitly by eq. \ref{singleroot}. Panel (b): $p = 0.95$. Thick black curves in the regions $z < z_{c,1}^{(bonds)}(p)$, eq. \ref{z1crit}, and $z > z_{c,2}^{(bonds)}(p)$, eq. \ref{z2crit},
are the single limit solution of  eq. \ref{p1}. Two dashed blue curves
in the intermediate region $z_{c,1}^{(bonds)}(p) < z < z_{c,2}^{(bonds)}(p)$ describe alternating state and are
defined by two distinct limit curves $x_{\rm odd}(p,z)$ and $x_{\rm even}(p,z)$, eqs. \ref{pp} and \ref{ppp}. The figure illustrates appearing of broken symmetry phase for $p>8/9$. Phase transitions between disordered and ordered state occur at $z_{c,1}^{(bonds)}(p)$ and $z_{c,2}^{(bonds)}(p)$, where blue lines meet with black lines.
}
\end{center}
\label{Fig4}
\end{figure}
Turning next to the case $p = 0.95$, we observe a more complicated behaviour.  First of all, we notice that similarly to the case $p \equiv 1$,
there is an apparent transition at a certain $z = z_{c,1}^{(bonds)}(p)$
from the sequence convergence to a single limit $x(p,z)$, to a subsequence convergence to alternating limits,
when the even and odd terms in the recursion in eq. \ref{rec3} converge, respectively, to some well-defined curves $x_{even}(p,z)$ and $x_{odd}(p,z)$. Curiously enough, however,
these two curves $x_{even}(p,z)$ and $x_{odd}(p,z)$ meet each other again at  some $z = z_{c,2}^{(bonds)}(p)$, such that for
$z \geq z_{c,2}^{(bonds)}(p)$ the sequence convergence to a single limit $x(p,z)$ is again restored.
While, evidently, the transition at $z = z_{c,1}^{(bonds)}(p)$ is the same as observed by Runnels;
that being, the transition from a disordered into an ordered phase, the second transition taking place at $z = z_{c,2}^{(bonds)}(p)$ can be identified as a re-entrant transition from the ordered into a disordered phase. To the best of our knowledge, we are unaware of any previous report on the presence of the latter re-entrant transition for a repulsive lattice gas on a Bethe lattice, although
it is quite plausible from the physical point of view. Indeed, for $p < 1$ the possibility of having two particles at neighbouring sites is not strictly prohibited and the penalty for having such a pair can be paid by an  increase of the chemical potential.
 Below we study the loci and the nature of these transitions in more details.

As in the previous section,
we reiterate the recursion in eq. \ref{rec3} to get a recursion scheme involving the terms of the same parity only:
\begin{eqnarray}
\label{rec4}
 x_N(p,z) &=& \left(1 + z (1 - p) \left(\frac{1 + z (1 - p) x_{N-2}^2(p,z)}{1 + z  x_{N-2}^2(p,z)}\right)^2\right) \nonumber\\
\fl &\times& \left(1 + z  \left(\frac{1 + z (1 - p) x_{N-2}^2(p,z)}{1 + z  x_{N-2}^2(p,z)}\right)^2\right)^{-1} \,,\,\,\, x_0(p,z) = 1\,.
\end{eqnarray}
Supposing next that $x_N(p,z) \to x(p,z)$ as $N \to \infty$, we find that $x(p,z)$ obeys the fifth-order equation which, again, conveniently factors into
\begin{eqnarray}
\label{p}
&&\Big(z x^3(p,z) - z (1- p) x^2(p,z) + x(p,z) - 1\Big) \times \nonumber\\
 &\times& \Big(z \left(1 + z (1-p)^2\right) x^2(p,z) - p z x(p,z) + 1 + z (1- p)   \Big) = 0 \,.
\end{eqnarray}
While the first factor is just our previous eq. \ref{p1}, which has one real root defining the solution in the single limit case, the second factor  has two solutions:
\begin{equation}
\label{pp}
x_{even}(p,z) = \frac{p z + \sqrt{D(p,z)}}{2 z \left(1 + z (1 - p)^2\right)} \,,
\end{equation}
and
\begin{equation}
\label{ppp}
x_{odd}(p,z) = \frac{p z - \sqrt{D(p,z)}}{2 z \left(1 + z (1 - p)^2\right)}\,,
\end{equation}
where the discriminant $D(p,z)$ is given by :
\begin{eqnarray}
\label{discr}
D(p,z) &=& - z \left(4 (1 - p)^3 z^2 + (8 - 12 p + 3p^2) z + 4\right) \nonumber\\
&=& 4 (1- p)^3 z \left(z - z_{c,1}^{(bonds)}(p)\right) \left(z_{c,2}^{(bonds)}(p) - z\right)
\end{eqnarray}
with
\begin{equation}
\label{z1crit}
z_{c,1}^{(bonds)}(p) = - \frac{8 - 12 p + 3 p^2 + p^{3/2} \sqrt{9 p - 8}}{8 (1 - p)^3} \,,
\end{equation}
and
\begin{equation}
\label{z2crit}
z_{c,2}^{(bonds)}(p) = \frac{- 8 + 12 p - 3 p^2 + p^{3/2} \sqrt{9 p - 8}}{8 (1 - p)^3} \,.
\end{equation}
We note now the following : both $z_{c,1}^{(bonds)}(p)$ and $z_{c,2}^{(bonds)}(p)$  in eqs. \ref{pp} and \ref{ppp} are real and positive only for $ p  >  p_c^{(bonds)} = 8/9$. The discriminant $D(p,z) > 0$ and hence, the solutions $x_{even}(p,z)$ and $x_{odd}(p,z)$, eqs. \ref{pp} and \ref{ppp}, are real and positive when $p > p_c^{(bonds)}$ and $z_{c,1}^{(bonds)}(p) < z < z_{c,2}^{(bonds)}(p)$,
so that a critical behaviour can emerge only for such values of $p$ and for such a range of $z$.
For $p < 8/9$ the second factor in  eq. \ref{p}  does not have real solutions and the only real root of eq. \ref{p} is determined by the
single limit solution, eq. \ref{singleroot}. This is precisely the behaviour we observe in Fig. 3 in which the panel (a) corresponds to the single-limit case with $p = 0.85 < p_c^{(bonds)}$, while the panel (b) with $p = 0.95 > p_c^{(bonds)}$
shows the subsequence convergence
to alternating limiting curves $x_{even}(p,z)$ and $x_{odd}(p,z)$ for $z_{c,1}^{(bonds)}(p) < z < z_{c,2}^{(bonds)}(p)$,
and the sequence convergence for $z < z_{c,1}^{(bonds)}(p)$ and
$z > z_{c,2}^{(bonds)}(p)$.  In
 Fig. 4 we depict
 the parametric curve which
solves the equation $D(p,z) = 0$. The region encircled by this
 curve corresponds to the critical region $z_{c,1}^{(bonds)}(p) < z < z_{c,2}^{(bonds)}(p)$.
When $p \to 1$, $z_{c,1}^{(bonds)}(p) \to 4$, (the Runnels' result), while $z_{c,2}^{(bonds)}(p) \sim 1/8 (1-p)^3 \to \infty$, which explains why the re-entrant transition is absent in the Runnels' case $p \equiv 1$.
We note lastly that $z_{c,1}^{(bonds)}(p)$ and $z_{c,2}^{(bonds)}(p)$ can be obtained from the stability 
analysis of the 
derivative of the function $f(x_{N-1}(p,z))$, which defines the recursion in eq. \ref{rec3}.
 Substituting into equation 
$f'(x_{N-1}(p,z))=-1$
 the solution \ref{pp} or \ref{ppp} instead of $x_{N-1}(p,z)$, and solving the resulting quadratic 
 equation with respect to $z$, one gets the values defined in eq.\ref{z1crit} and eq. \ref{z2crit}.

\begin{figure}
\begin{center}
\includegraphics[width=.48\hsize]{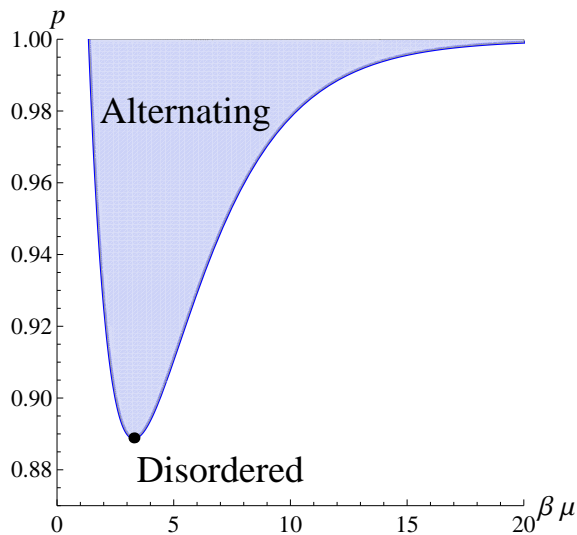}
\caption{Phase diagram for the annealed Model I. Blue line separating disordered and ordered (alternating) phases is the solution of equation $D(p,z) = 0$ with $D(p,z)$ defined by eq. \ref{discr}. The full circle indicates the location of the critical value of the parameter $p = p_c^{(bonds)} = 8/9$.
}
\end{center}
\label{Fig5}
\end{figure}

\subsection{Order parameter for $p < 1$}

The order parameter $\delta \rho$ is defined as the difference of average occupations of adjacent layers $j$ and $j+1$ in the
limit $j \to \infty$. For $p < 1$, the average density is still described by eq. \ref{density} with $t_j(z)$ replaced by $t_j(p,z)$, which obeys the recursion
\begin{equation}
\label{rrr}
 t_j(p,z) =  b_j(p,z) \left(1 - 2 \frac{p \, \left(1 - b_{j-1}(p,z)\right)}{1 - p \, b_{j-1}(p,z)} t_{j-1}(p,z) \right) \,, \,\,\, t_0(p,z) =1 \,,
\end{equation}
where $b_j(p,z) = (1 - x_j(p,z))/p$.
The recursion in eq. \ref{rrr} can be solved exactly to give
\begin{equation}
\label{tjjj}
t_j(p,z) = \sum_{i = 0}^{j-1} (-2)^j \prod_{n=j-i}^{j} b_n(p,z) \left(\frac{p \left(1 - b_n(p,z)\right)}{1 - p \, b_n(p,z)}\right)^{1 - \delta_{n,j}} \,,
\end{equation}
where $\delta_{n,j}$ is the Kronecker-delta, such that $\delta_{n,n} = 1$ and is zero, otherwise.
Further on, inserting eq. \ref{tjjj} into eq. \ref{density} and re-arranging the series, we get the expansion in eq. \ref{exp} with $r_j(z)$ replaced by $r_j(p,z)$,
which now obeys the recursion relation
\begin{equation}
\label{gen}
r_j(p,z) = b_j(p,z) \left(1 - \frac{p \, \left(1 - b_j(p,z)\right)}{1 - p \, b_j(p,z)} \, r_{j+1}(p,z)\right) \,,
\end{equation}
whose solution is given by
\begin{equation}
\label{zz}
r_j(p,z) = \sum_{i=0}^{\infty} (-1)^{i} \prod_{n=j}^{i+j} b_n(p,z) \left(\frac{p \, \left(1 - b_n(p,z)\right)}{1 - p \, b_n(p,z)}\right)^{1 - \delta_{n,i+j}} \,.
\end{equation}
Similarly to the  expression in eq. \ref{exp} for the $p \equiv1$ case, eq. \ref{zz} defines the average occupation at generation $j$ in the general case when $0 \leq p \leq 1$.

\begin{figure}
\begin{center}
\includegraphics[width=.6\hsize]{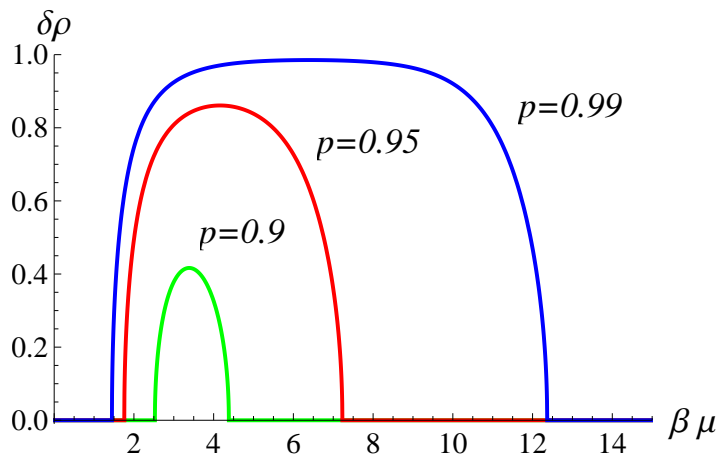}
\caption{Order parameter $\delta \rho$ in eq. \ref{parameter} versus $\beta \mu$ for different $p > p_c^{(bonds)}$. Green curve corresponds to $p=0.9$, the red one - to $p=0.95$
and the blue curve - to $p = 0.99$.
}
\end{center}
\label{Fig6}
\end{figure}

We focus now on the single limit case obtained either for $p < p_c^{(bonds)} =8/9$, or for $p > p_c^{(bonds)}$ with $z < z_{c,1}^{(bonds)}(p)$ or $z > z_{c,2}^{(bonds)}(p)$. In this case all $b_j(p,z)$ converge as $j \to \infty$ to $b(p,z)= (1 - x(p,z))p^{-1}$ with $x(p,z)$ defined by eq. \ref{singleroot}. In virtue of eq. \ref{gen}, we have
that the average occupation at generation $j$ converges to
\begin{equation}
\label{rpz}
r(p,z) = \frac{b(p,z) \left(1 - p \, b(p,z)\right)}{1 - p \, b^2(p,z)} \,,
\end{equation}
such that the order parameter $\delta \rho \equiv 0$. Further on, we consider the critical case when $p > p_c^{(bonds)}$ and $z \in [z_{c,1}^{(bonds)}(p),z_{c,2}^{(bonds)}(p)]$. Here, $b_j(p,z)$ with even $j$ converges in the limit $j \to \infty$ to $b_{-}(p,z) = (1 - x_{even}(p,z))p^{-1}$, eq. \ref{pp}, while $b_j(p,z)$ with odd $j$ converges to $b_{+}(p,z) = (1 - x_{odd}(p,z))p^{-1}$, eq. \ref{ppp}. Correspondingly, the average occupations on the generation $j$ deep in the interior of an infinitely large
tree are given
by
\begin{equation}
\label{--}
r_{-}(p,z) = \frac{b_-(p,z) \left(1 - p \, b_+(p,z)\right)}{1 - p \, b_+(p,z) b_-(p,z)}
\end{equation}
for even $j$, and
\begin{equation}
\label{++}
 r_{+}(p,z) = \frac{b_+(p,z) \left(1 - p \, b_-(p,z)\right)}{1 - p \, b_+(p,z) b_-(p,z)}
\end{equation}
for odd $j$, respectively. In consequence, the staggered density obeys
\begin{eqnarray}
\label{parameter}
 \delta \rho &=&  r_{+}(p,z) -  r_{-}(p,z) \nonumber\\
\fl &=& \frac{b_+(p,z) - b_-(p,z)}{1 - p \, b_+(p,z) b_-(p,z)} \nonumber\\
\fl &=& \frac{x_{even}(p,z) - x_{odd}(p,z)}{1 - p + p \, \left(x_{even}(p,z) + x_{odd}(p,z) -  x_{even}(p,z) x_{odd}(p,z)\right)} \,,
\end{eqnarray}
where $x_{even}(p,z)$ and $x_{odd}(p,z)$ are defined by eqs. \ref{pp} and \ref{ppp}. In Fig. 5 we depict the behaviour of the order parameter in eq. \ref{parameter} for several values of $p > p_c^{(bonds)}$.

Using eqs. \ref{pp} to \ref{discr}, we can formally 
rewrite eq. \ref{parameter} in the form
\begin{equation}\label{delrz}
\delta \rho=\frac{\sqrt{4 (1-p)^3 z\left(z-z^{(bonds)}_{c,1}(p)\right)\left(z^{(bonds)}_{c,2}(p)-z\right)}}{(1-p)^3 z^2+(1-2 (1-p) p) z-p}.
\end{equation}
For $z$ sufficiently close to $z^{(bonds)}_{c,1}(p)$ one has
\begin{equation}\label{cvz1}
\fl z=\exp(\ln(z^{(bonds)}_{c,1}(p))/(1-\tau) )\simeq z^{(bonds)}_{c,1}(p) + z^{(bonds)}_{c,1}(p) \ln(z^{(bonds)}_{c,1}(p)) \tau,
\end{equation}
where $\tau=(T-T_c)/T_c$ is the deviation of the reduced temperature 
from the critical point $T_c$. Then, substituting eq. \ref{cvz1}
 into eq. \ref{delrz} and expanding the resulting expression in powers of $\tau$, we find
\begin{equation}\label{delrz}
\fl \delta \rho \simeq\frac{\sqrt{4 (1-p)^3 z^{(bonds)}_{c,1}(p)\left(z^{(bonds)}_{c,2}(p)-z^{(bonds)}_{c,1}(p)\right)\ln(z^{(bonds)}_{c,1}(p))}}{(1-p)^3 {z^{(bonds)}_{c,1}(p)}^2+(1-2 (1-p) p) z^{(bonds)}_{c,1}(p)-p}\sqrt{\tau}+O(\tau^{3/2}),
\end{equation} 
which implies that   
the order parameter has the form  $\delta\rho\sim \tau^b$ with the mean-field critical exponent $b=1/2$. 
In a similar way, we analyse the behaviour of the order parameter in the vicinity of the inverted transition point; that being, 
$z = z^{(bonds)}_{c,2}(p)$,  
to get a scaling behaviour with the same critical exponent $b=1/2$.

\subsection{Mean densities and the compressibility}

We address next the question of the order of the transitions which our system undergoes at $z = z_{c,1}^{(bonds)}(p)$ and at $z = z_{c,2}^{(bonds)}$ for $p > p_c^{(bonds)}$.
Consider first the behaviour of the mean particle density at these points. In the sequence convergence limit, i.e., for $z < z_{c,1}^{(bonds)}(p)$ or for $z > z_{c,2}^{(bonds)}(p)$,
the average occupation of the generations with odd and even $j$ deep in the interior of the infinite tree is the same, such that  in virtue of eqs. \ref{exp} and \ref{rpz} we have
\begin{equation}
\label{density1}
\rho_{out}(p,z) = r(p,z) =  \frac{b(p,z) \left(1 - p \, b(p,z)\right)}{1 - p \, b^2(p,z)} \,,
\end{equation}
where the subscript "out" signifies that we deal with the behaviour out of the critical region,
$b(p,z) = (1 - x(p,z))/p$
and $x(p,z)$ is given by eq. \ref{singleroot}.

Further on, within the critical region $z \in [z_{c,1}^{(bonds)}(p),z_{c,2}^{(bonds)}(p)]$, the average occupations of all the sites at even and odd generations are defined by eqs. \ref{--} and \ref{++}, respectively, such that the mean density within the critical region obeys
\begin{eqnarray}
\label{density2}
\rho_{in}(p,z) &=& \frac{1}{2} \left( r_{+}(p,z) + r_{-}(p,z)\right)  \nonumber\\
&=& \frac{1}{2} \frac{b_-(p,z) + b_+(p,z) - 2 p b_+(p,z) b_-(p,z)}{1 - p b_+(p,z) b_-(p,z)} \nonumber\\
&=& \frac{2 + (2 - 3 p) z}{2 + 2 z \left(2 - 3 p + (1 - p)^3 z\right)} \,.
\end{eqnarray}

Next, one may readily notice that since
\begin{equation}
b(p,z_{c,1}^{(bonds)}(p)) = b_-(p,z_{c,1}^{(bonds)}(p)) = b_+(p,z_{c,1}^{(bonds)}(p))
\end{equation}
 and
\begin{equation}
 b(p,z_{c,2}^{(bonds)(p)}) = b_-(p,z_{c,2}^{(bonds)}(p)) = b_+(p,z_{c,2}^{bonds)}(p)),
\end{equation}
one has $\rho_{out}(p,z_{c,1}^{(bonds)}(p)) = \rho_{in}(p,z_{c,1}^{(bonds)}(p))$ and $\rho_{out}(p,z_{c,2}^{(bonds)}(p)) = \rho_{in}(p,z_{c,2}^{(bonds)}(p))$, such that
the density is a piece-wise continuous function of $z$.

\begin{figure}
\begin{center}
\includegraphics[width=.6\hsize]{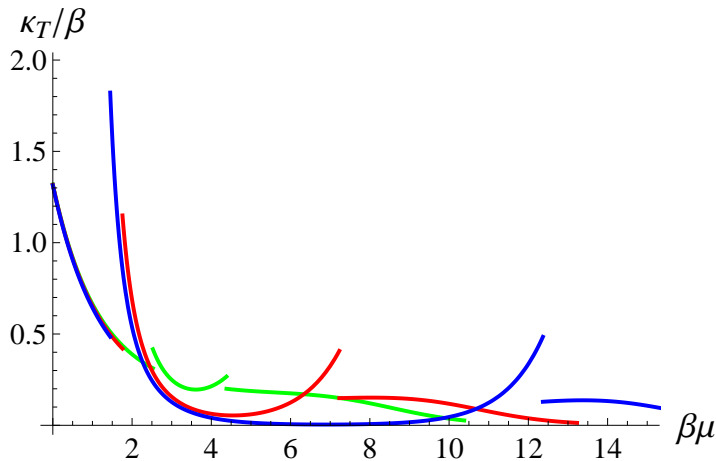}
\caption{Compressibility $\kappa_T/ \beta$ in eqs. \ref{comp1} and \ref{comp2} versus $\beta \mu$ for different $p > p_c^{(bonds)}$. Green curve corresponds to $p=0.9$, the red one - to $p=0.95$
and the blue curve - to $p=0.99$.
}
\end{center}
\label{Fig7}
\end{figure}

Consider next the compressibility $\kappa_T$ defined as
\begin{equation}
\label{kappa_def}
\kappa_T = \frac{1}{\rho^2} \frac{\partial \rho}{\partial \mu} = \frac{\beta}{\rho^2} \frac{\partial \rho}{\partial \ln(z)} \,.
\end{equation}
Away of the critical region, i.e., for $z < z_{c,1}^{(bonds)}(p)$ or for $z > z_{c,2}^{(bonds)}(p)$, we use the expression in eq. \ref{density1} to get
\begin{eqnarray}
\label{comp1}
\kappa_T/\beta &=&  \frac{\bigg(x(p,z) - 1 + p\bigg)}{x(p,z) \bigg(1 - x(p,z)\bigg)} \frac{\Big(p \left(1 - 2 x(p,t)\right) - \left(1 - x(p,z)\right)^2\Big)}{\Big(p \left( x(p,t) - 2\right) + 2 \left(1 - x(p,z)\right)^2\Big)} \,.
\end{eqnarray}
On the other hand, within the critical region, i.e., for $z \in [z_{c,1}^{(bonds)}(p),z_{c,2}^{(bonds)}(p)]$, we take advantage of the expression in eq. \ref{density2} to get
\begin{equation}
\label{comp2}
\kappa_T/\beta = \frac{2 z \Big( 3p - 2 - 4  (1-p)^3 z - (2 - 3 p) (1-p)^3 z^2\Big)}{\Big(2 + (2 - 3 p) z\Big)^2} \,.
\end{equation}
In Fig. 6 we plot the compressibility $\kappa_T$ defined by eqs. \ref{comp1} and \ref{comp2} versus $\beta \mu$ for different values of $p > p_c^{(bonds)}$. We observe that the compressibility exhibits a finite upward jump when the system enters into the ordered phase at $z = z_{c,1}^{(bonds)}(p)$, and also a finite downward jump when it re-enters the disordered phase for $z$ exceeding $z_{c,2}^{(bonds)}(p)$. This implies that both transitions are continuous, or second order in the Ehrenfest nomenclature. In particular, for $p = 0.9$,   $\kappa/\beta \to 5/16$ when $z$ approaches $z_{c,1}^{(bonds)}(p)$ from below, and $\kappa/\beta \to 5/12$ when $z \to z^{(bonds)}_{c,1}(p)$ from above. For the re-entrant transition for the same value of $p$, we have $\kappa/\beta \to 1/5$ for $z \to z_{c,2}^{(bonds)}(p)$ from below, and   $\kappa/\beta \to 4/15$ for $z \to z_{c,2}^{(bonds)}(p)$ from above, respectively.

We close this Section with the following remark:
Exact calculation of
the average occupation at generation $j$ in the general case $0 \leq p \leq 1$, culminating at our eq. \ref{zz},  which
we performed here by generalising very directly the Runnel's
approach, appears to be quite cumbersome even for the simple model of a purely repulsive lattice gas (Model I)
 and thus we can hardly expect that an analogous analysis can be carried out for the  Model II with its multi-site Boolean interactions.
In this regard, it might be instructive to present a somewhat simpler derivation of the main results of this Section using a different approach, which may also shed some light on the physical meaning
 of the expressions
  for $\rho_{in}(z)$ in eq. \ref{density2} and for $\rho_{out}(z)$ in eq. \ref{density1}.
 Let $\rho_0$ denote the mean density  at the central site  of a finite Cayley tree. This density is formally defined as
 \begin{equation}\label{rho_b}
\rho_0=\frac{ Z^{(bonds,1)}_N(p)}{Z^{(bonds,0)}_N(p)+Z^{(bonds,1)}_N(p)},
\end{equation}
where $Z^{(bonds,1)}_N(p)$ and $Z^{(bonds,0)}_N(p)$ are given by eqs. \ref{b0}.
Taking into account the definition of $x_N(p,z)$ in
eq. \ref{xx}, we arrive at the following expression:
\begin{equation}\label{rho_b0}
\rho_0=\frac{z x^3_N(p,z)}{1+z x_N^3(z,p)} \,.
\end{equation}
Outside of the critical region,  in the limit $N\to\infty$,  mean densities on each site of the tree are equal to the same quantity $\rho_{out}$, and consequently, $\rho_{out}=\rho_0$. Plugging  $x_N(z,p)\to x(z,p)$ into eq. \ref{rho_b0} and combining it with eq. \ref{p1}, we recover the expression in
eq. \ref{density1}. Further on, within the critical region the situation is a bit more delicate,
because here we have two alternating limits for $N\to \infty$: $x_{even}(p,z)$ and $x_{odd}(p,z)$.
Let $\rho_{even}(p,z)$ and $\rho_{odd}(p,z)$  obey
\begin{equation}\label{g1}
\rho_{even}(p,z)=\frac{z x^3_{even}}{1+z x_{even}^3}  \,, \rho_{odd}(p,z)=\frac{z x^3_{odd}}{1+z x_{odd}^3}  \,.
\end{equation}
Defining next the mean density in the critical region
as $\rho_{in}=(\rho_{even}+\rho_{odd})/2$, and using  eqs. \ref{pp}
 and \ref{ppp}, we obtain for $\rho_{in}$ exactly
the same expression as the one in eq. \ref{density2}.

\section{Model II : Catalytic sites}
\label{m2}

We turn next to the annealed version of the Model II and examine the critical behaviour of the annealed partition function defined by our eqs. \ref{deff} and \ref{bool}, as the function
of the reaction probability $p$ and of the chemical potential $\mu$ (or the activity $z$).
Similarly to the previously considered case of catalytic bonds,
we decompose formally
$Z^{(\rm sites)}(p)$
in eq. \ref{deff}
as
$Z^{(\rm sites)}(p) = Z_N^{(\rm sites,0)}(p) + Z_N^{(\rm sites,1)}(p) $, where
$Z_N^{(\rm sites,0)}(p)$ and $ Z_N^{(\rm sites,1)}(p)$ denote the
disorder-averaged  partition functions of the Cayley tree with a vacant and an occupied central sites.
One immediately notices that the  partition function of the entire tree with a vacant central site factorises into
the product of the partition functions defined on the subtrees; that being,
we still have
$Z_N^{(\rm sites,0)}(p) = G_N^3(0;p)$,
where $G_N(0,p)$ is the partition function in eq. \ref{deff} defined on a subtree (see, Fig. \ref{Fig1}, right panel) with a vacant root. Note that, evidently, $G_N(0;p=1)=B_N(0;p=1) = B_N(0)$.

The case of a tree with an occupied central site is more difficult, compared
to the Model I, since here different subtrees emanating from the occupied central node are effectively coupled via the Boolean function $\Psi_0$, eq. \ref{bool}, and the way how they may get decoupled depends now on the occupations of the sites belonging to the first generation.
In consequence,
we have to represent $Z_N^{(\rm sites,1)}(p)$ as the sum of all configurations with different occupations of the sites of the first generation; namely, all configurations  with an occupied central site and a) all three sites in the first generation occupied, b) two sites occupied and one - vacant, c) one site occupied and two - vacant, and c) all three sites in the first generation vacant. Introducing next auxiliary functions $G(n_0,n_1;p)$, which denote the partition function of a subtree with the root having an occupation number $n_0$ and the site at the first generation having the occupation number $n_1$, we represent $Z_N^{(\rm sites,1)}(p)$ by directly counting all possible configurations as
\begin{eqnarray}
\label{occupied}
Z_N^{(\rm sites,1)}(p) &=& \frac{G_N^3(1,1;p)}{z^2 (1 - p)^2} + 3 \frac{G_N^2(1,1;p) G_N(1,0;p)}{z^2 (1 - p)} \nonumber\\
&+& \frac{3 G_N(1,1;p) G_N^2(1,0;p)}{z^2} + \frac{G_N^3(1,0;p)}{z^2} \,,
\end{eqnarray}
 Further on, we find straightforwardly that $G(n_0,n_1;p)$ obey the following recursions:
\begin{eqnarray}
\label{gg}
G_N(0,0;p) &=& G_{N - 1}^2(0;p) \,, \,\,\,
G_N(1,0;p) = z G_{N-1}^2(0;p) \,, \nonumber\\
G_N(0,1;p) &=& \frac{1}{(1-p) z} \Big((1-p) G_{N-1}^2(1,0;p) + \nonumber\\
 &+& 2 (1 - p) G_{N-1}(1,0;p) G_{N-1}(1,1;p) + G_{N-1}^2(1,1;p)\Big) \,, \nonumber\\
G_N(1,1;p) &=& \Big((1-p) G_{N-1}(1,0;p) + G_{N-1}(1,1,p)\Big)^2 \,.
\end{eqnarray}
Recalling next that $G_N(1;p) = G_N(1,1;p) + G_N(1,0;p)$ and $G_N(0;p) = G_N(0,0;p) + G_N(0,1;p)$, we find from eq. \ref{gg} the recursions obeyed by $G_N(1;p)$ and $G_N(0;p)$:
\begin{eqnarray}
\label{0}
G_N(0;p) &=& G_{N-1}^2(0;p) + p z G_{N-2}^4(0;p) + \nonumber\\
&+& \frac{1}{(1-p) z} \Big(G_{N-1}(1;p) - p z G_{N-2}^2(0;p)\Big)^2
\end{eqnarray}
and
\begin{equation}
\label{1}
G_N(1;p) = z G^2_{N-1}(0;p) + \Big(G_{N-1}(1;p) - p z G_{N-2}^2(0;p)\Big)^2 \,.
\end{equation}
Now, it is convenient to introduce
new auxiliary
 functions
\begin{equation}
\label{defin}
\psi_N(p,z)= \frac{G_N(1;p)}{z G_{N-1}^2(0;p)} \,; \xi_N(p,z) = \frac{G^2_{N-1}(0;p)}{G_{N}(0;p)} \,,
\end{equation}
which obey, in virtue of eqs. \ref{0} and \ref{1}, the following
coupled recursion relations:
\begin{eqnarray}
\label{ggg}
\psi_N(p,z) &=& 1 + z \xi^2_{N-1}(p,z) \Big(\psi_{N-1}(p,z) - p\Big)^2 \,, \psi_0(p,z) = 1 \,, \nonumber\\
\xi_N(p,z) &=& \left(1 + p z \xi^2_{N-1}(p,z) + z \xi^2_{N-1}(p,z)  \frac{\left(\psi_{N-1}(p,z) - p\right)^2}{1-p}\right)^{-1}  \,,
\end{eqnarray}
where $\xi_0(p,z) = 1$.
Note, that the parameter $x_N(p,z)$ used to simplify the recursion in the Model I is given now by $x_N(p,z) = \psi_N(p,z) \xi_N(p,z)$. Note, as well, that in the limit $p \to 1$ all $\psi_N \to 1$, such that the second equation in eqs. \ref{ggg} becomes $\xi_N = (1 + z \xi^2_{N-1})^{-1}$, which is just our previous eq. \ref{rec1} written in term of the function $\xi_N(p,z)$.

We turn back now to our eq. \ref{occupied}.
Using the representations
\begin{eqnarray}
G_N(1,1;p) &=& G_N(1;p) - G_N(1,0;p) \nonumber\\
&=& z G_{N-1}^2(0;p) \Big(\psi_N(p,z) - 1\Big) \nonumber\\
&=& z \xi_N(p,z) G_N(0;p)  \Big(\psi_N(p,z) - 1\Big) \,,
\end{eqnarray}
as well as our eqs. \ref{gg}, we may rewrite eq. \ref{occupied} formally
as
\begin{equation}
\label{zzzz}
 Z_N^{(\rm sites,1)}(p) = \frac{z \xi_N^3(p,z) G_N^3(0;p)}{(1- p )^2} \Big(\bigg(\psi_N(p,z)  - p \bigg)^3 + p (1-p)^2 \Big) \,,
\end{equation}
which together with the relation $Z_N^{(\rm sites,0)}(p) = G_N^3(0;p)$ define the  partition function $Z^{(\rm sites)}(p)$ of the entire Cayley tree.

Finally, interpreting the definition of the function $\xi_N(p,z)$ in eq. \ref{defin} as a recursion relation with an evident "initial" condition $G_0(0;p) = 1$, we find that $G_N(0;p)$ for an arbitrary $N$ can be expressed via $\xi_N(p,z)$ as
\begin{equation}
G_N(0;p) = \prod_{j=1}^N \xi_j^{-2^{N-j}}(p,z) \,,
\end{equation}
such that eq. \ref{pressure2} yields for the pressure of the adsorbate
\begin{eqnarray}
\label{pressure3}
P_{II}(T,\mu) &=& \frac{1}{\beta} \lim_{M \to \infty} \frac{\ln Z^{(\rm sites)}(p)}{M} = \frac{3}{\beta} \lim_{M \to \infty} \frac{\ln G_N(0;p)}{M}  \nonumber\\
&+&  \frac{1}{\beta} \lim_{M \to \infty} \frac{1}{M} \ln \left[ 1+ \frac{z \xi_N^3(p,z)}{(1- p )^2} \left(\left(\psi_N(p,z) -p \right)^3 + p (1-p)^2 \right) \right] \nonumber\\
&=& \frac{1}{\beta} \sum_{j=1}^{\infty} 2^{-j} \ln\left(\frac{1}{\xi_j(p,z)}\right) \,,
\end{eqnarray}
where the sum in the last line defines the desired thermodynamic limit result for the entire Cayley tree. In \ref{A2} we  explain how one may subtract the contribution due to the boundary sites and get an analogous expression for the deep interior of the Cayley tree - the Bethe lattice.

\subsection{Solution of the recursion relations for $p < 1$}

Before we proceed further, it seems again
expedient to get first some general understanding of the behaviour of the recursion scheme
defined by eqs. \ref{ggg}. To this end, we
 generate first $20$ terms of $\psi_N(p,z)$ and of $\xi_N(p,z)$  for three different values of the parameter $p$: $p=0.6$, $p = 0.8$ and $p = 0.9$. These results are plotted in Figs. 7 and 8. We notice that the situation appears to be somewhat similar to the one encountered for the annealed Model I in the sense that also here
 two clearly distinct behaviours are observed : for $p = 0.6$ we have a convergence to the single limit curve for all values of $z$, while
 for larger $p$
 there is a
 range of $z$ with an apparent single limit convergence, and a bounded region in which
 odd and even terms seemingly
 converge to alternating curves.
This signifies that, first, there exists some critical value $p_{c,1}^{(sites)}$ of the parameter $p$, which
lies somewhere in-between $p=0.6$ and $p=0.8$.
Second, it shows that also for the annealed Model II for $p > p_{c,1}^{(sites)}$ there are some
critical value $z_{c,1}^{(sites)}(p)$, at which the systems enters from a disordered phase into an ordered one,  and some critical value
$z_{c,2}^{(sites)}(p) > z_{c,1}^{(sites)}(p)$, at which the systems re-enters into a disordered phase.

There is, however, a notable distinction between the behaviour of $\psi_N(p,z)$ and $\xi_N(p,z)$ for $p = 0.8$ and $p = 0.9$:
while for $z$ close to $z_{c,1}^{(sites)}(p)$ we observe quite a similar smooth behaviour, in the vicinity of $z_{c,2}^{(sites)}(p)$ the functions  $\psi_N(p,z)$ and
$\xi_N(p,z)$ approach the single limit curve much more abruptly for $p = 0.9$ than for $p = 0.8$. This
hints that for the former case we may encounter a phase transition of a different type.

\begin{figure}
\begin{center}
\includegraphics[width=.48\hsize]{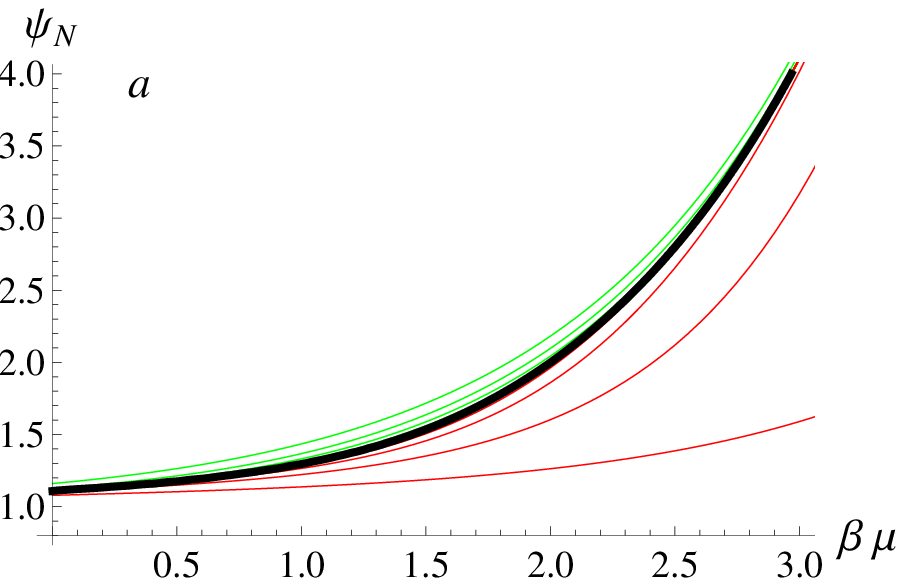}
\includegraphics[width=.48\hsize]{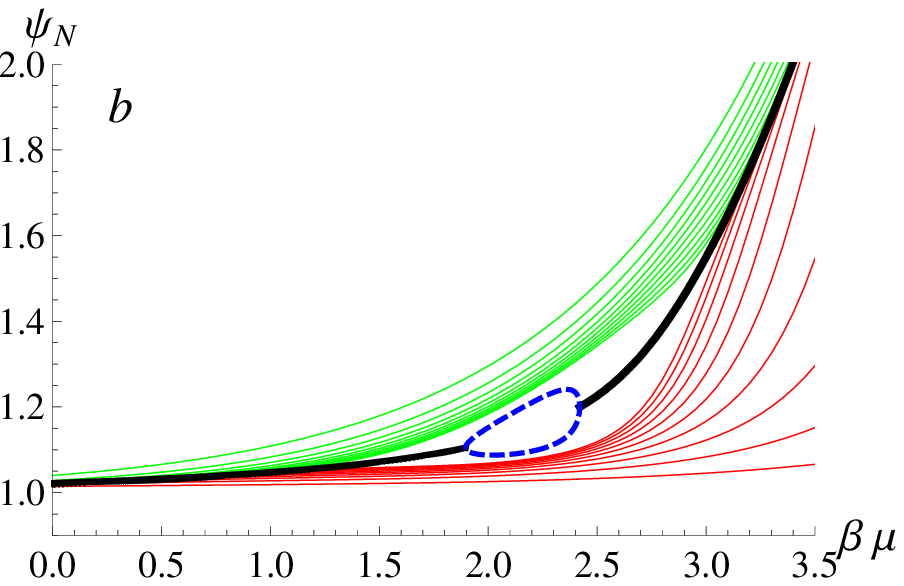}
\includegraphics[width=.48\hsize]{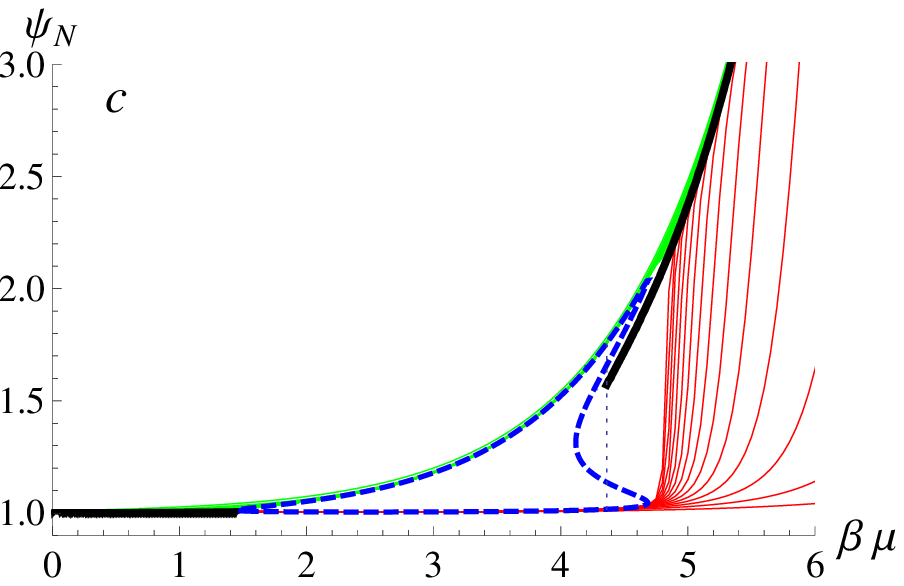}
\caption{$\psi_N=\psi_N(p,z)$ in eqs. \ref{ggg} versus $\beta \mu$ for $N=1,2,3,\ldots,20$.
Green curves correspond to odd $N$, while the red ones - to even values of $N$. Panel (a) : $p = 0.6$, Panel (b): $p = 0.8$ and Panel (c): $p = 0.9$.
Thick black lines are
defined by eq. \ref{psi_via_xi} with $\psi(p,t)$ being the single limit solution of eq. \ref{kk}. Thick blue dashed lines defining the solution in the alternating limits
are the roots of eq. \ref{rootsf2}. Vertical thin dotted line in Panel (c) shows the discontinuity in $\psi_N(p,z)$ at $z = z_{c,2}^{(sites)}(p)$ emerging for $p>p_{c,2}^{(sites)}$, where $p_{c,2}^{(sites)}$ is calculated from eq. \ref{cond}. }
\end{center}
\label{Fig8}
\end{figure}

\begin{figure}
\begin{center}
\includegraphics[width=.47\hsize]{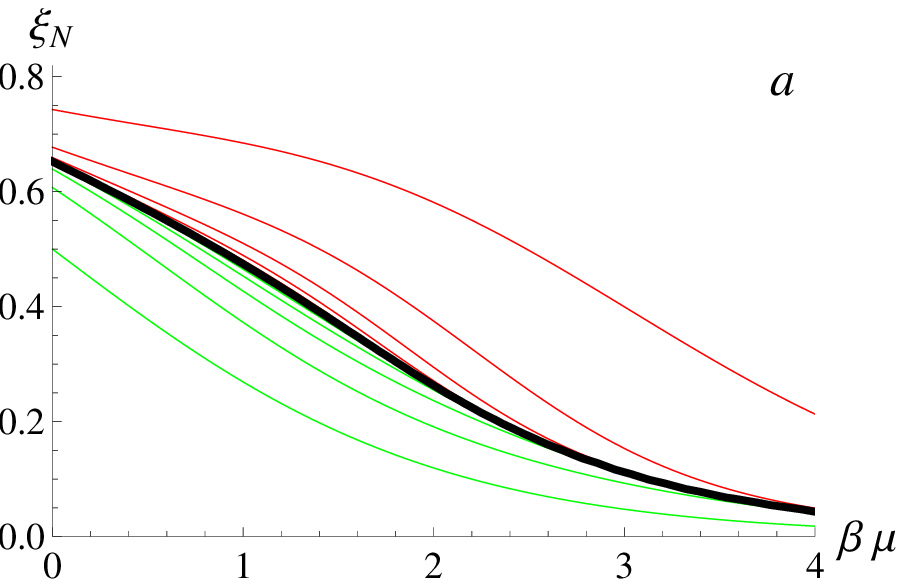}
\includegraphics[width=.47\hsize]{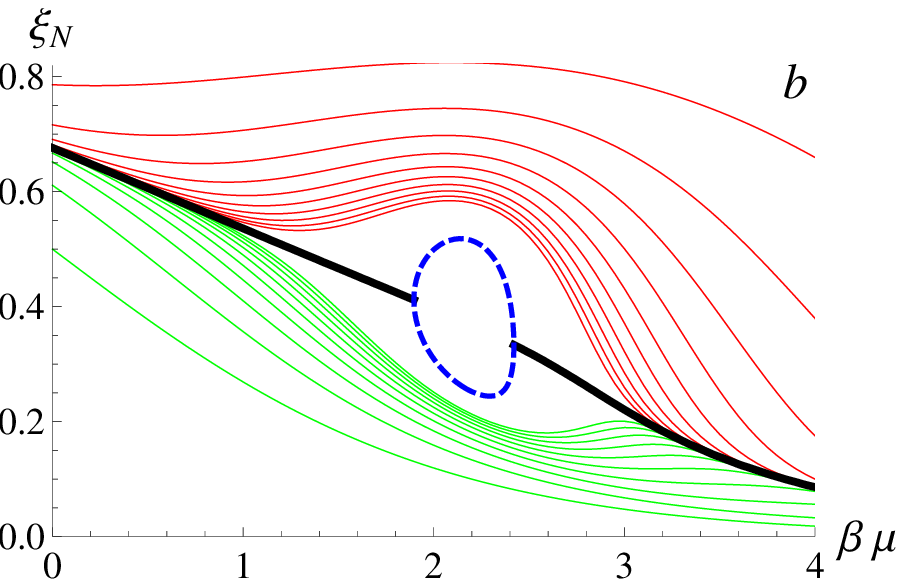}
\includegraphics[width=.47\hsize]{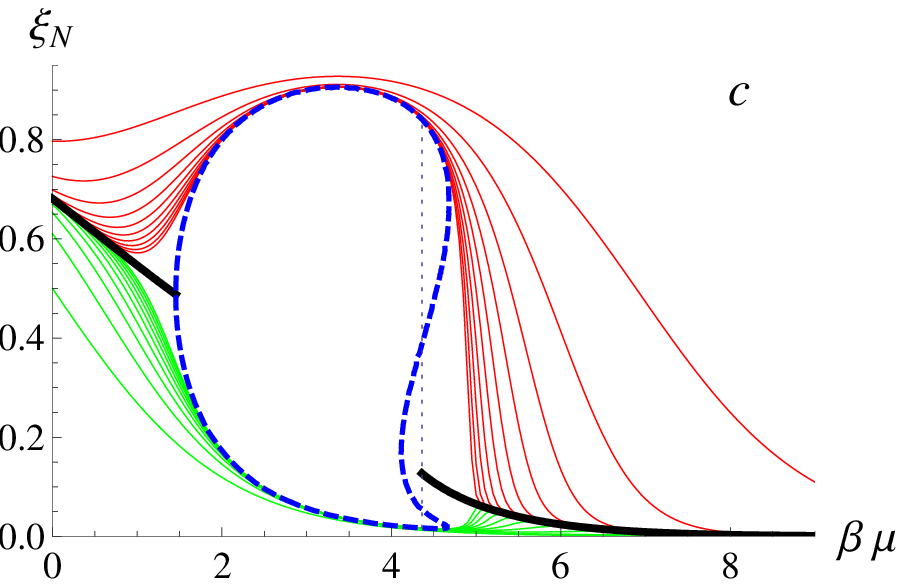}
\caption{$\xi_N=\xi_N(p,z)$ in eqs. \ref{ggg} versus $\beta \mu$ for $N=1,2,3,\ldots,20$.
Green curves correspond to odd $N$ while the red ones - to even values of $N$. Panel (a) : $p = 0.6$, Panel (b): $p = 0.8$ and Panel (c): $p = 0.9$.
Thick black lines  are the single limit solutions,  eq. \ref{kk}. Thick blue dashed lines define the alternating limits solution of eq. \ref{rootsf2} expressed in terms of $\xi(p,z)$. Vertical thin dotted line in Panel (c)
shows the discontinuity in $\xi(p,z)$ at $z = z_{c,2}^{(sites)}(p)$ for $p > p_{c,2}^{(sites)}$;  $p_{c,2}^{(sites)}$ is obtained from eq. \ref{cond}.}
\end{center}
\label{Fig9}
\end{figure}


We turn first to the single limit case. Supposing that  $\xi_N(p,z) \to \xi(p,z)$ and $\psi_N(p,z) \to \psi(p,z)$ as $N \to \infty$,
we have that eqs. \ref{ggg} become
 \begin{eqnarray}
\label{gggg}
\psi(p,z) &=& 1 + z \xi(p,z)^2 \Big(\psi(p,z) - p\Big)^2 \,,  \nonumber\\
\xi(p,z) &=& \left(1 + p z \xi^2(p,z) + z \xi(p,z)^2  \frac{\bigg(\psi(p,z) - p\bigg)^2}{1-p}\right)^{-1}  \,.
\end{eqnarray}
Further on, it follows from eq. \ref{gggg} that
\begin{equation}\label{psi_via_xi}
\psi(p,z) - p = (1 - p) \left(\frac{1}{\xi(p,z)} - p z \xi^2(p,z)\right) \,,
\end{equation}
which permits us to write down  a closed-form equation determining $\xi$ :
\begin{eqnarray}
\label{kk}
f_1(\xi(p,z),p,z)&=&p^2 (1- p) \xi^7(p,z) z^3 - 2 p (1-p) \xi^4(p,z) z^2 + \nonumber\\
&+& \xi(p,z) \left(p \xi^2(p,z) + 1 - p\right) z + \xi(p,z) - 1 = 0 \,,
\end{eqnarray}
which is a seventh-order equation in $\xi(p,z)$. Note that for $p \equiv 1$,  $\psi(p,z) \equiv 1$ such that
eq. \ref{kk} reduces to a depressed cubic equation of the form $z \xi(p,z)^3 + \xi(p,z) - 1 = 0$, which is just our previous eq. \ref{small}.
For general $p < 1$, we can only solve equation eq. \ref{kk} numerically. The real roots $\xi(p,z)$, eq. \ref{kk}, for fixed $p=0.6,\,0.8,\,0.9$ are depicted by a thick black line in Fig. 8, (and correspondingly, the roots of $\psi_N(p,z)$ are depicted in Fig. 7),
demonstrating a convergence of the recursion
to this single limit curve for any $z$ for $p = 0.6$, as well as for $z < z_{c,1}^{(sites)}(p)$ and $z > z_{c,2}^{(sites)}(p)$ for $p=0.8$ and $p=0.9$,
(where $z_{c,1}^{(sites)}(p)$ and $z_{c,2}^{(sites)}(p)$ stay undefined, for the moment).

Consider now the alternating limits case. As in the case of the Model I, (and also following the Runnels' analysis \cite{runnels}), we reiterate eqs. \ref{ggg} once more
to get a recursion involving only the terms of the same parity with respect to $N$. This gives
\begin{eqnarray}
\label{g5}
\psi_N(p,z) &= & 1 + \frac{z (1-p)^2 K_{N-2}^2(p,z)}{\bigg(p (1-p) z \xi_{N-2}^2(p,z) + K_{N-2}(p,z) \bigg)^2} \,, \nonumber\\
\xi_N(p,z) &=& \left(1 + \frac{z (1 - p) \bigg(p (1-p) + K_{N-2}^2(p,z)\bigg)}{\bigg(p(1-p) z \xi^2_{N-2}(p,z) + K_{N-2}(p,z)\bigg)^2}\right)^{-1} \,,
\end{eqnarray}
where we denote
\begin{equation}
\label{g50}
K_{N-2}(p,z) = 1 - p + z \xi_{N-2}^2(p,z) \Big(\psi_{N-2}(p,z) - p\Big)^2 \,.
\end{equation}
Further on, we assume that $\xi_N (p,z)\to \xi(p,z)$, $\psi_N(p,z) \to \psi(p,z)$ and $K_N(p,z) \to K(p,z)$ as $N \to \infty$, so that we may rewrite eqs. \ref{g5} and \ref{g50} as:
\begin{eqnarray}
\label{g6a}
\psi(p,z) &= & 1 + \frac{z (1-p)^2 K^2(p,z)}{\left(p (1-p) z \xi^2(p,z) + K(p,z) \right)^2} \,, \\ \label{g6b}
\xi(p,z) &=& \left(1 + \frac{z (1 - p) \left(p (1-p) + K^2(p,z)\right)}{\left(p(1-p) z \xi^2(p,z) + K(p,z)\right)^2}\right)^{-1}
\end{eqnarray}
where
\begin{equation}
\label{K}
K(p,z) = 1 - p + z \xi^2(p,z) \Big(\psi(p,z) - p\Big)^2 \,.
\end{equation}
As a matter of fact,  from our eq. \ref{K} we can express $\psi(p,z)$ in eq. \ref{g6a} via $K(p,z)$, which will result in a closed system of equations for $K(p,z)$ and $\xi(p,z)$. Next, expressing $K(p,z)$ from the second equation via $\xi(p,z)$ and plugging the result into the first equation, we find a general closed-form equation for $\xi(p,z)$. Some straightforward (but rather tedious) calculations, which we omit here, show that the closed-form equation for $\xi(p,z)$ can be cast into the form
\begin{equation}
\label{f1f2}
f_1(\xi(p,z), p,z)  f_2(\xi(p,z), p,z) = 0 \,,
\end{equation}
where function $f_1(\xi(p,z), p,z)$ is defined in eq. \ref{kk}. The function $f_2(\xi(p,z), p,z)$ is a polynomial of the fourteenth order in $\xi(p,z)$ and a polynomial of the eighth order in $z$, and is presented in an explicit form in \ref{A} (see eq. \ref{f2}).

\begin{figure}
\begin{center}
\includegraphics[width=.48\hsize]{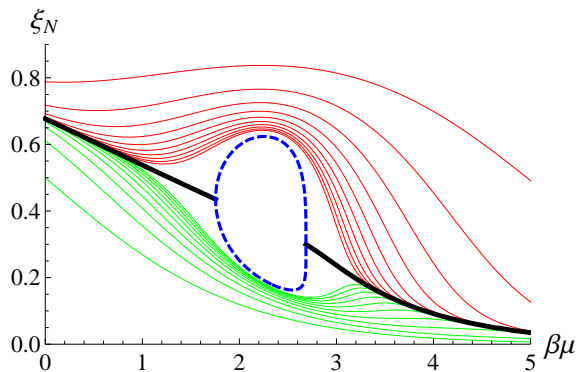}
\caption{$\xi_N=\xi_N(p,z)$ in eq. \ref{ggg} versus $\beta \mu$ for $N=1,2,3,\ldots,20$ in the critical case  $p = p_{c,2}^{(sites)}\backsimeq 0.813$.
Green curves correspond to odd $N$, while the red ones - to even values of $N$. Thick black lines  are the single limit solutions  of eq. \ref{kk} while the thick dashed lines are the alternating limits solutions of eqs. \ref{rootsf2}. Note that $\xi_N(p,z)$ becomes multi-valued at $z = z_{c,2}^{(sites)}(p)$. }
\end{center}
\label{Fig10}
\end{figure}

Now, the root of the $f_1(\xi(p,z), p,z) = 0$, eq. \ref{kk}, defines the solution in the single limit case. On the other hand, similarly to the situation described in Sec. \ref{m1},  the roots of the equation
\begin{equation}
\label{rootsf2}
f_2(\xi(p,z),p,z) = 0 \,,
\end{equation}
define the solutions in the alternating limits case. Numerical analysis of a strongly non-linear eq. \ref{rootsf2} together
with some complementary arguments presented in \ref{B}
 permit us to draw the following conclusions:
\begin{itemize}
\item Two alternating limit solutions appear only for $p>p_{c,1}^{(sites)} \backsimeq 0.794$, which value is specific, of course, to the coordination number three of the Bethe lattice.
\item the ordered phase can only exist for $p>p_{c,1}^{(sites)}$ and such $z$ which obey the double-sided inequality $z_{c,1}^{(sites)}(p) < z < z_{c,2}^{(sites)}(p)$.
\item There is another critical value of the reaction probability $p=p_{c,2}^{(sites)}\backsimeq 0.813$, such
that for $p_{c,1}^{(sites)}<p<p_{c,2}^{(sites)}$  two alternating
limit curves meet each other at the single limit solution, eq. \ref{kk}, at the points $z_{c,1}^{(sites)}(p)$ and $z_{c,2}^{(sites)}(p)$.
\item For $p>p_{c,2}^{(sites)}$ we have a different behaviour: the alternating
limit curves have different values at $z = z_{c,2}^{(sites)}(p)$ and the alternating limit solution of eqs. \ref{rootsf2} become multi-valued and no longer
represents the actual limits approached by $\psi_N(p,z)$ and $\xi_N(p,z)$ as $N \to \infty$ (see Figs. 7 and 8, and also Fig. 9 which depicts the behaviour for the critical value $p = p_{c,2}^{(sites)}$ in which case the alternating limit solutions of eqs. \ref{rootsf2} form a vertical line at $z = z_{c,2}^{(sites)}(p)$).
This signifies that both $\psi_N(p,z)$ and $\xi_N(p,z)$ exhibit a \textit{discontinuous} transition at $z = z_{c,2}^{(sites)}(p)$ for $N = \infty$.
\end{itemize}
These findings are summarised in Fig. 10, in which we depict
the dependence of $z_{c,1}^{(sites)}(p)$ and $z_{c,2}^{(sites)}(p)$ on $p$  defining the complete phase diagram for the annealed Model II. Note that similarly to the Model I, $z_{c,2}^{(sites)}(p) \to \infty$ when $p \to 1$, such that the re-entrant transition disappears, as it should.

\begin{figure}
\begin{center}
\includegraphics[width=.48\hsize]{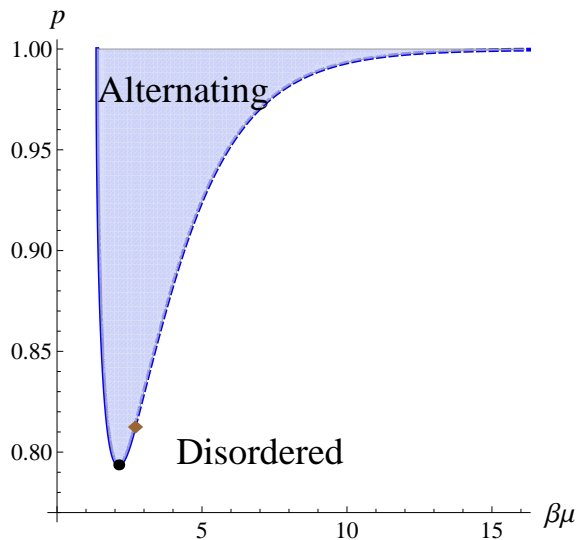}
\caption{Phase diagram for the annealed Model II. Solid line demarcates a continuous phase transition between ordered (alternating) and disordered phases, while the dashed curve shows the discontinuous re-entrant transition between these two phases.
The full circle  corresponds to $p_{c,1}^{(sites)}$, while the full diamond shows the location of the tricritical point, at which the lines of continuous and discontinuous transitions meet.
}
\end{center}
\label{phase}
\end{figure}

\subsection{Mean density, compressibility and the order parameter for $p < 1$ }

We start with the calculations of the mean density and of the compressibility in the single limit case, i.e., outside of the critical region.
The mean density  at the central site of the Cayley tree is formally defined as
 \begin{equation}\label{rho_nodes}
\rho_0=\frac{ Z^{(\rm sites,1)}_N(p)}{Z^{(\rm sites)}(p)}=\frac{ Z^{(\rm sites,1)}_N(p)}{Z^{(\rm sites,0)}_N(p)+Z^{(\rm sites,1)}_N(p)},
\end{equation}
where
 $ Z^{(\rm sites,1)}_N(p)$ is defined by eq. \ref{zzzz} and $Z_N^{(\rm sites,0)}(p) = G_N^3(0;p)$. This  implies that the
 mean density at the central site of a Cayley tree with $N$ generations is given by
 \begin{equation}\label{rho_nodes2}
\rho_0=\frac{{z \xi^3_N(p,z)}[(\psi_N(p,z)-p)^3+p(1-p)^2]}{{(1-p)^2}+{z \xi^3_N(p,z)}[(\psi_N(p,z)-p)^3+p(1-p)^2]}.
\end{equation}
For a deep interior of an infinite Cayley tree, i.e., for the Bethe lattice, and outside of the critical region, that is, for any $z$ and $p < p_{c,1}^{(sites)}$, or  for an arbitrary $p$ but $z$ which are either
 less than $z_{c,1}^{(sites)}(p)$ or greater than $z_{c,2}^{(sites)}(p)$, mean densities at each site (including the central one) are equal. Consequently, supposing that for the Bethe lattice
  $\psi_N(p,z) = \psi(p,z)$ and $\xi_N(p,z) = \xi(p,z)$, and using  eq. \ref{psi_via_xi}, we obtain
 \begin{equation}\label{rho_first}
 \rho_{out}^{(sites)}(p,z) =\frac{{z }\left[(1-p)(1-p z\xi^3(p,z))^3+p \xi^3(p,z)\right]}{1+{z }\left[(1-p)(1-p z\xi^3(p,z))^3+p \xi^3(p,z)\right]} \,.
\end{equation}
Compressibility in this case can be found from the definition in eq. \ref{kappa_def}. Substituting eq. \ref{rho_first} into
eq. \ref{kappa_def},
we find that the compressibility away of the critical region obeys
\begin{eqnarray}
\label{compressibility}
\fl \kappa_T/\beta&=& \frac{1}{{z }\Big((1-p)(1-p z\xi^3(p,z))^3+p \xi(p,z)\Big)^3}+\nonumber\\
\fl &+& \frac{{3 p  \xi^2(p,z)}\Big((1-p)\left(1-p z \xi^3(p,z)\right)\left(3z \xi'(p,z)+ \xi(p,z)\right)+ \xi'(p,z)\Big)}{\left(\left((1-p)(1-p z\xi^3(p,z))^3+p \xi(p,z)\right)^3\right)^2} \,,
\end{eqnarray}
where the derivative $\xi'(p,z) = \partial \xi(p,z)/\partial z$ is defined  as:
\begin{equation}
 \frac{\partial \xi(p,z)}{\partial z}=-\frac{\partial f_1(\xi(p,z),p,z)}{\partial z} \left(\frac{\partial f_1(\xi(p,z),p,z)}{\partial \xi(p,z)}\right)^{-1} \,.
\end{equation}

\begin{figure}
\begin{center}
\includegraphics[width=.48\hsize]{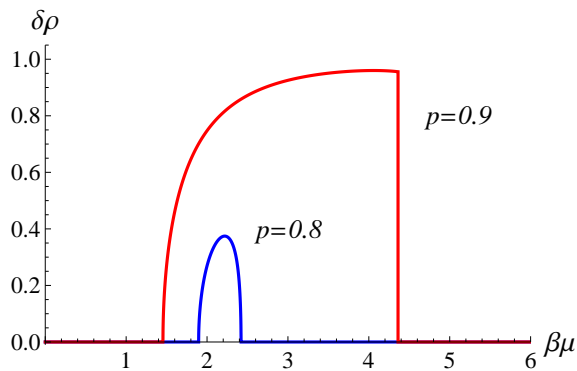}
\caption{Order parameter $\delta \rho$  versus $\beta \mu$ for different $p > p_{c,1}$. Blue curve corresponds to $p=0.8$, the red one - to $p=0.9$.
}
\end{center}
\label{orderp}
\end{figure}

We turn next to the behaviour in the critical region. We note that here, in contrast to the Model I,
the exact calculation of the densities in the alternating layers is hardly possible due to a very complicated form of the recursion schemes obeyed by $\xi_N(p,z)$ and $\psi_N(p,z)$.
Therefore, we resort here to an approach described at the end of the previous Section.
Within the critical region, i.e., for $p>p_{c,1}^{(sites)}$ and for $z_{c,1}^{(sites)}(p)<z<z_{c,2}^{(sites)}(p)$,  mean densities of even and odd generations are different. Consequently, in this case we
may
consider mean density $\rho_{in}^{(sites)}(p,z)=(\rho_{odd}^{(sites)}(p,z)+\rho_{even}^{(sites)}(p,z))/2$.  Using eq. \ref{rho_nodes2}, we then define the
mean density of sites at the odd generation, $\rho_{odd}^{(sites)}(p,z)$, and at the even generation, $\rho_{even}^{(sites)}(p,z)$,  as
\begin{equation}\label{rho_plus}
 \rho_{odd}^{(sites)}(p,z)=\frac{{z \xi_{odd}^3(p,z)}\Big[(\psi_{odd}(p,z)-p)^3+p(1-p)^2\Big]}{{(1-p)^2}+{z \xi_{odd}^3(p,z)}\Big[(\psi_{odd}(p,z)-p)^3+p(1-p)^2\Big]},
\end{equation}
and
\begin{equation}\label{rho_minus}
 \rho_{even}^{(sites)}(p,z)=\frac{{z \xi_{even}^3(p,z)}\Big[(\psi_{even}(p,z)-p)^3+p(1-p)^2\Big]}{{(1-p)^2}+{z \xi_{even}^3(p,z)}\Big[(\psi_{even}(p,z)-p)^3+p(1-p)^2\Big]} \,.
\end{equation}
Using eqs. \ref{rho_plus} and \ref{rho_minus}, we can define the order parameter
$\delta \rho=|\rho_{odd}-\rho_{even}|$. The latter  is depicted  in Fig. 11 as the function of $\beta \mu$ for the case  $p=0.8$ (i.e., for $p$ within the interval
 $p_{c,1}^{(sites)} < p < p_{c,2}^{(sites)}$)
and for the case $p=0.9$ (i.e., for $p$ such that $p > p_{c,2}^{(sites)}$).
One observes a markedly different behaviour at $z = z_{c,2}^{(sites)}(p)$ - in the former case $\delta \rho$ attains a maximal value for $z < z_{c,2}^{(sites)}(p)$ and smoothly approaches zero, while in the latter case the order parameter vanishes \textit{discontinuously} at $z = z_{c,2}^{(sites)}(p)$.

The compressibility in this case can be found if we replace
 $\rho$ in eq. \ref{kappa_def} by $\rho_{in}$. Then, the only properties  we need to know
are the derivatives of $\xi_{odd}(p,z)$,  $\xi_{even}(p,z)$, $\psi_{odd}(p,z)$ and  $\psi_{even}(p,z)$ with respect to $z$.
The latter
can be calculated rather straightforwardly and we obtain the following result:
\begin{equation}\label{QQ}
\kappa_T/\beta=2 z \frac{A_1 A'_2 -A_2  A'_1}{\Big(A_1+2 A_2\Big)^2} \,,
\end{equation}
 where the explicit expressions for
 the functions $A_1= A_1(s(p,z),\sigma(p,z))$ and $A_2= A_2(s(p,z),\sigma(p,z))$ are given by eqs. \ref{a11} and \ref{a22}  (see \ref{B}), respectively.
 Their derivatives with respect to $z$, i.e., $A'_1= A'_1(s(p,z),\sigma(p,z))$ and $A'_2= A'_2(s(p,z),\sigma(p,z))$
 are also presented in an explicit form
 in  \ref{B} (see eqs. \ref{derivs} and \ref{derivsigma}).

\begin{figure}
\begin{center}
\includegraphics[width=.48\hsize]{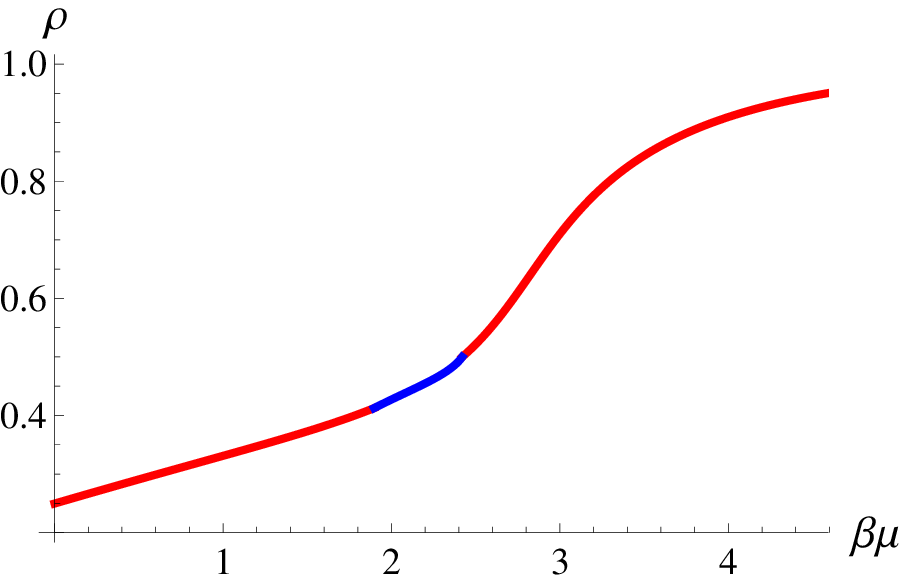}
\includegraphics[width=.48\hsize]{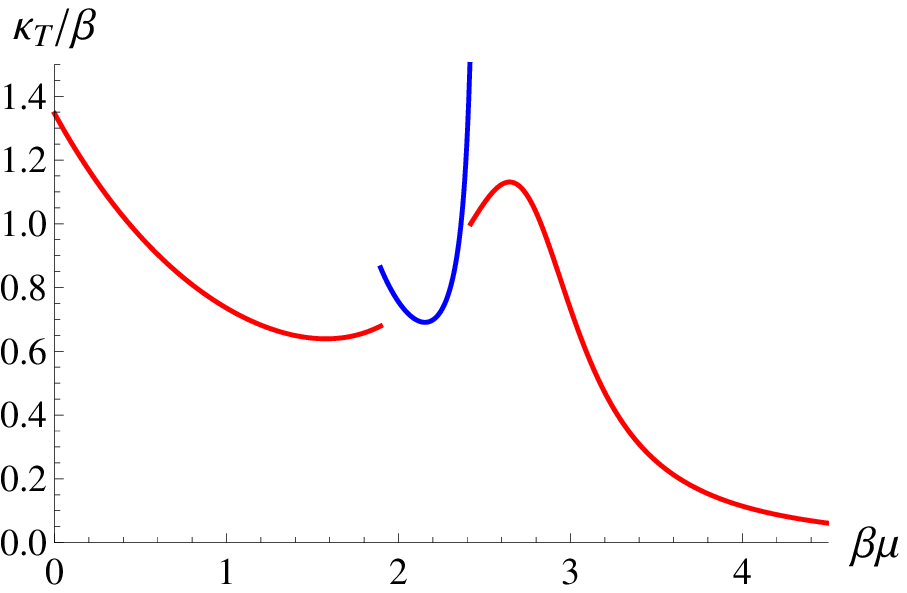}
\caption{Mean density $\rho$, panel (a), and the compressibility $\kappa_T/ \beta$, panel (b), versus $\beta \mu$ for $p=0.8$.
The red curves in panels (a) and (b) correspond to eqs. \ref{rho_first} and \ref{compressibility}, respectively.
The blue curve in the middle of the panel (a) depicts $\rho_{in}=(\rho_{odd}^{(sites)}(p,z)+\rho_{even}^{(sites)}(p,z))/2$ with $\rho_{odd}^{(sites)}(p,z)$ and $\rho_{odd}^{(sites)}(p,z)$
defined by  eqs. \ref{rho_minus} and \ref{rho_plus}. The blue curve in the panel (b)  shows the compressibility in the critical region defined by eq. \ref{QQ}.
}
\end{center}
\label{compr}
\end{figure}

\begin{figure}
\begin{center}
\includegraphics[width=.48\hsize]{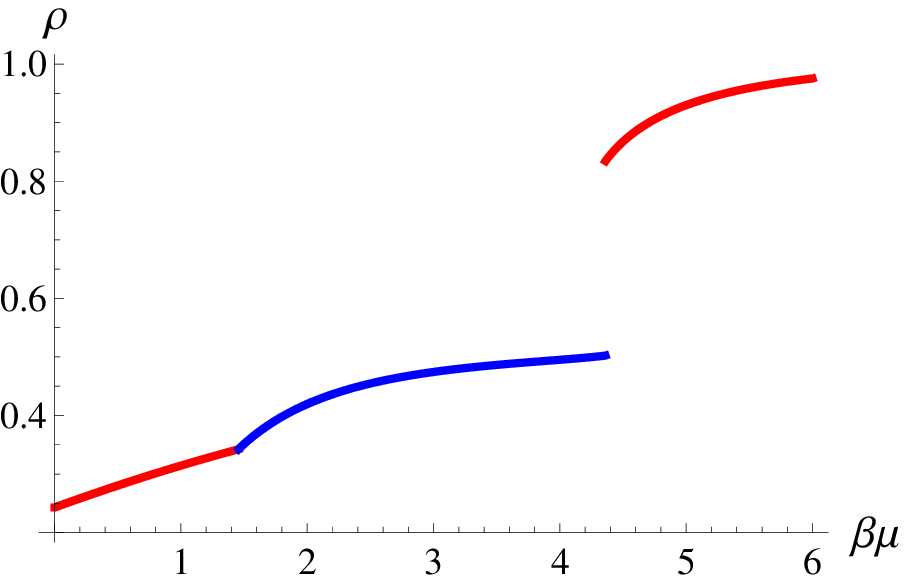}
\includegraphics[width=.48\hsize]{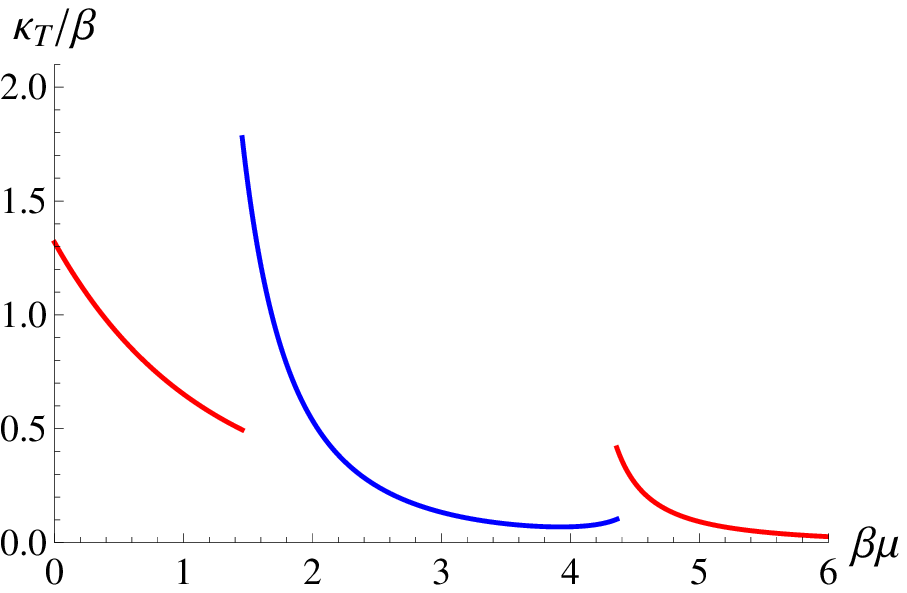}
\end{center}
\caption{Mean density $\rho$, panel (a), and the compressibility $\kappa_T/ \beta$, panel (b),  versus $\beta \mu$ for $p=0.9$. The red curves in panels (a) and (b) correspond to eqs. \ref{rho_first} and \ref{compressibility}, respectively.
The blue curve in the middle of the panel (a) depicts $\rho_{in}=(\rho_{odd}^{(sites)}(p,z)+\rho_{even}^{(sites)}(p,z))/2$ with $\rho_{odd}^{(sites)}(p,z)$ and $\rho_{odd}^{(sites)}(p,z)$
defined by  eqs. \ref{rho_minus} and \ref{rho_plus}. The blue curve in the panel (b) defines the compressibility in the critical region defined by eq. \ref{QQ}.
}
\label{compr2}
\end{figure}

Therefore, we obtain the compressibility for the whole range of $\beta \mu$ and arbitrary $p$. The results for $p=0.8$ and $p=0.9$ are presented in Figs. 12 and 13, respectively.  One notices that for $p=0.8$ the density is piece-wise \textit{continuous}
with cusps at $z  = z_{c,1}^{(sites)}(p)$ and at $z = z_{c,2}^{(sites)}(p)$), while
the compressibility exhibits finite jumps when the system enters into the ordered phase at $z = z_{c,1}^{(sites)}(p)$ and  re-enters into the disordered phase  at  $z=z_{c,2}^{(sites)}(p)$. This implies that both transitions are \textit{continuous} transitions.
On the other hand,
for $p = 0.9$,  (which value of $p$  exceeds $p_{c,2}^{(sites)}$), we observe that the
density is piece-wise continuous at $z = z_{c,1}^{(sites)}(p)$, and varies \textit{discontinuously} at $z= z_{c,2}^{(sites)}(p)$, which signifies that in this range of values of the reaction probability $p$
the re-entrant transition is
\textit{discontinuous}. Interestingly enough, behaviour of the compressibility
after the re-entrant transition is markedly different depending whether the latter
is continuous or discontinuous. In case of a continuous
 transition the compressibility jumps \textit{downwards} at the transition point,
 passes through a maximum and then decreases  monotonically with $z$.
 In case of a discontinuous transition
 the compressibility makes an \textit{upward} jump and then monotonically decreases with $z$.

\section{\label{conc} Conclusions}

To conclude, we have studied here  equilibrium
properties of two lattice-gas models
of catalytically-activated
$A + A \to \oslash$ reactions on a lattice of adsorption sites. In both models,
the $A$ particles are assumed to be in thermal contact with their vapour phase (a reservoir maintained at a
chemical potential $\mu$), adsorb onto empty adsorption sites and desorb from the lattice.
We considered two different ways of modelling
such reactions: in the Model I we assumed that some fraction $p$ of the \textit{bonds} connecting
neighbouring adsorption sites possesses special catalytic properties so that any two $A$s appearing on the sites connected by such a bond react instantaneously and desorb. In the Model II, we stipulated that some fraction $p$ of the adsorption \textit{sites} possesses such properties and
the reaction  takes place once
at least one of the neighbouring $A$ particles occupies a catalytic adsorption site.

We
focused on the case of annealed disorder in the distribution of the catalytic bonds or sites,
which is tantamount to the situation when the reaction between two $A$s may take place at any point on the lattice but happens with a finite probability $p$, which means that the reaction is not perfect (instantaneous) but is characterised by some finite reaction constant.
In this case the Model I describes a lattice gas with soft, purely
repulsive nearest-neighbour interactions, while the Model II represents a lattice gas with particular
multi-site interactions of particles: here,
the interaction of an adsorbed particle
with its nearest environment has a Boolean form - it
is either zero, in case when the particle does not have any neighbour, or is a constant independent of the actual
number of neighbouring adsorbed particles, if at least one of them is present.
We provided exact analytical solutions for the annealed versions of both Models I and II on the interior of the Cayley tree - the Bethe lattice,  and showed that they exhibit
 a rich "critical" behaviour with respect to $\mu$ and $p$, characterised by a transition into an ordered state, which for both models is continuous with a finite jump in compressibility,
and a re-entrant transition into a disordered phase, which is continuous with a finite jump in compressibility for the Model I and, depending on the value of $p$, may be either continuous or  discontinuous with a finite jump of density, for the Model II.

The wealth of critical phenomena which we observed  for the Models I and II  for a Bethe lattice geometry, certainly merits further, more deep investigation including a study of the kinetic behaviour in both models, as well  
as an analysis of both models on more realistic, regular or random adsorbent lattices
with quenched (above and below the percolation threshold) and/or annealed distributions of the catalysts. 
Lastly, we would like to remark that it is interesting to extend our analysis over the case of a
monomer-monomer model $A + B \to \oslash$, which involves two types of particles  \cite{ziff2,doug,alb,redner}.
This model with a finite reactivity $p$ has been already 
studied for one-dimensional finite
systems in Ref. \cite{pop} and also for
two-dimensional lattices with an \textit{annealed} distribution of the catalytic bonds \cite{pop1,pop2}. In Refs. \cite{pop1,pop2} it was shown that, in particular,  the grand canonical partition function  for the $A + B \to \oslash$ model
can be mapped onto the partition function of the
general spin $S =1$ model \cite{blume,blume1}, which permitted to
exploit the large number of results available for the latter (see, e.g., Ref. \cite{pop2}). For instance, for the symmetric case of equal chemical potentials for both species, a phase transition was predicted from the phase in which 
the particles of both sorts have the same densities, to the phase in which one sort of particles prevails.  This transition can be of the first order or a continuous one, depending on the precise values of the system's parameters. It was also shown in Ref. \cite{pop1}
that in some parameter space the monomer-monomer model defined on a honeycomb lattice with an annealed disorder 
in placement of the catalytic bonds,
reduces to the original
Blume-Emery-Griffiths model \cite{BEG},
whose solution can be obtained in a closed form
via a mapping to a zero-field Ising model on a regular honeycomb
lattice, which can be solved exactly and exhibits a symmetry-breaking continuous transition \cite{hor2,wu}.
For the
$A + B \to \oslash$ reaction on a lattice with catalytic sites no analytical results are available at present.

\subsection*{Acknowledgments}

The authors
wish to thank Professor S. Dietrich for fruitful discussions.
MD  acknowledges the hospitality of the Max-Planck-Institute Stuttgart, where the most
 of work on this project has been done.
OB is partially supported by the
European Research Council  Grant
No. FPTOpt-277998. MD is supported in part by  the FP7 EU IRSES
project No. 612707 ``Dynamics of and in Complex Systems".

 \appendix

 \section{\label{A} Polynomial $f_2(\xi(p,z),p,z)$}

We present  here the polynomial $f_2(\xi(p,z),p,z)$ which enters eq. \ref{f1f2}:

\begin{eqnarray}
\label{f2}
\fl f_2(\xi(p,z), p,z)&=& 1+\xi  z \Big(\xi ^{13} q^3 p^4 z^6-\xi ^{12} q^3 p^4 z^6-2 \xi ^{10} q^4
   p^3 z^6\nonumber\\
 \fl  &+& 2 \xi ^{11} q^3 p^3 z^5 (q z{+}1)+\xi ^7 q^2 p^2 z^3 (3-q z (5- 8 q
   z))\nonumber\\
\fl   &+&\xi ^5 q p z^2 \left(z \left(16 p -7 p^2+q^4 z^2-11 q^3 z-9\right)+3\right)\nonumber\\
 \fl &+&\xi ^8 q^2 p^2 z^4 \left(7-q^3 z^2-10 q^2 z-10 p\right)\nonumber\\
 \fl &+&\xi ^9
   q^2 p^2 z^4 \left(q^4 z^3+3 q^3 z^2-(4 p-3) q z+2 p+1\right)\nonumber\\
\fl   &+&\xi ^4 q p z^2
   \left(q^3 z^2+16 q^2 z-10 p+7\right)-\xi ^2 q^2 p z^2\nonumber\\
   \fl &+&\xi  \left(2 q^3 z^2-(5
   p-4) q z-p+2\right)\nonumber\\
   \fl &-&\xi ^6 q^3 p z^3 \left(z \left(2 q^3 z^2-2 q^2 z+13
   p-10\right)-6\right)\nonumber\\
   \fl &+&\xi ^3 q z \left(q^5 z^4{+}4 q^4 z^3{-}(13 p{-}6) q^2 z^2{+}(9 p{-}4)
   q z{-}3 p{-}1\right){-}p\Big) \,,
\end{eqnarray}
where $q = 1 - p$.

\section{\label{A2} Pressure for the Bethe lattice}

To derive an explicit expression for the
substrate pressure of a deep interior of the Cayley tree - the Bethe lattice, we use the procedure which is well described in Ref. \cite{ananikian}.

\subsection{Pressure for the annealed Model I}

The procedure described in Ref. \cite{ananikian}
consists of the following steps. First,  using eq. \ref{pressure0} we can write the pressure of an adsorbate on an $N$-generation Cayley tree in the form :
\begin{equation}\label{pn}
\beta M P_{I,N}=\ln Z^{(bonds)} (p) \,.
\end{equation}
Using next the relation  $Z^{(bonds)} (p)=Z_N^{(bonds,0)} (p)+Z_N^{(bonds,1)} (p)$, as well as the definitions in eqs. \ref{b0} and \ref{xx}, we find that the pressure in eq. \ref{pn} reads
\begin{equation}
\beta M P_{I,N}=3\ln B_N(0,p)+\ln (1+z x_N^3(p,z)) \,.
\end{equation}
Further on, for $B_N(0,p)$ we use the relation in eq. \ref{333}, which gives
\begin{eqnarray}
\beta M P_{I,N}&=&6 \ln B_{N-1}(0,p)+3\ln(1+z x_{N-1}^2(p,z)) \nonumber\\
&+&\ln (1+z x_N^3(p,z)) \,.
\end{eqnarray}
Rewriting the latter expression in terms of the pressure $P_{I,N-1}$ for an $N-1$-generation Cayley tree we therefore arrive at
\begin{eqnarray}
\beta M P_{I,N}&=&2 \beta M P_{I,N-1}-2\ln (1+z x_{N-1}^3(p,z)) \nonumber\\
&+&3\ln(1+z x_{N-1}^2(p,z))+\ln (1+z x_N^3(p,z)) \,.
\end{eqnarray}
Reiterating this procedure $n$ times, we find the following
recursion relation for $P_{I,N}$ :
\begin{equation}
\beta M P_{I,N}=2^n \beta M P_{I,N-n}-\beta M P_{I,Nn} \,,
\end{equation}
where $P_{I,Nn}$ is now the
pressure of an $n$-generation deep interior of the Cayley tree - an $n$-generation Bethe lattice, which reads
\begin{eqnarray}
\beta M P_{I,Nn}&=&3\sum_{k=1}^n \ln(1+z x_{N-k}^2(p,z))-2^n\ln (1+z x_{N-n}^3(p,z))  \nonumber\\
&+&\ln (1+z x_N^3(p,z)) \,.
\end{eqnarray}
Lastly, we determine $P_{I}$ by dividing the above expression by $\beta M$, and taking next the limit $N\to\infty$, (in which limit all $x_{N-k}(p,z)\equiv x(p,z)$),  which eventually yieldes the desired expression
\begin{equation}\label{pi}
P_{I}(x(p,z))=\frac{3}{2\beta}\ln(1+z x^2(p,z))-\frac{1}{2\beta}\ln (1+z x^3(p,z)) \,.
\end{equation}

We note next that the mean density can be obtained from eq. \ref{pi}
in the standard way by differentiating the pressure with respect to $z$, i.e.,
v$\rho(z)=\beta z \partial P_{I}/\partial z$,  and
excluding $\partial x(p,z)/\partial z$  and $z$
from the resulting expression using
eq. \ref{p1}. In doing so, one arrives at the expression given in eq. \ref{density1}.
Pressure in the critical region, in which two alternating limits for $x_N(p,z)$ exist,
is to be written in the form
\begin{equation}
P_{I,in}=\frac{1}{2}\left(P_{I}(x_{even}(p,z))+P_{I}(x_{odd}(p,z))\right) \,.
\end{equation}
The mean density in this case is derived
 using eqs. \ref{pp} and \ref{ppp}, to get eventually
 the expression in eq. \ref{density2}.

\subsection{Pressure for the annealed Model II}
Applying the analogous procedure for the derivation of
the pressure for the Model II, we get following expression
\begin{eqnarray}\label{pii}
\fl P_{II}(\xi(p,z),\psi(p,z))&=&\frac{3}{2\beta}\ln \left(1+p z \xi^2(p,z) + \frac{z \xi^2(p,z)}{(1-p)}\bigg(\psi(p,z)-p\bigg)^2\right)
\nonumber\\ \fl &-&  \frac{1}{2\beta}\ln \left(1+\frac{z \xi^3(p,z)}{(1-p)^2}\bigg(\bigg(\psi(p,z)-p\bigg)^3+p(1-p)^2\bigg)\right) \,.
\end{eqnarray}
Requiring the continuity of the pressure
at the phase transition point, we determine the critical value of the activity $z_f(p)$ for the
 first order phase transition:
\begin{eqnarray}\label{cond}
P_{II}(\xi(p,z_f),\psi(p,z_f))&=&\frac{1}{2}\Big(P_{II}(\xi_{even}(p,z_f),\psi_{even}(p,z_f))+\nonumber\\
&+&P_{II}(\xi_{odd}(p,z_f),\psi_{odd}(p,z_f))\Big)  \,.
\end{eqnarray}
Together
 with eq. \ref{gggg} and eqs. \ref{gggg1} to \ref{gggg4},
 the expression in eq. \ref{cond} implicitly determines $z_f(p)$. The solution is depicted
   by a dashed line in Fig. 10.

\section{\label{B} Alternating limits solution for the annealed Model II}

Here we present some complementary analysis of the alternating limits solution for the annealed Model II,
which also turns out to be useful
for the calculations of the staggered density and of the compressibility.
We first
revisit our eq. \ref{g5} and suppose that
$\xi_N(p,z)$ and $\psi_N(p,z)$  with odd and even $N$  converge as $N \to \infty$ to $\xi_{odd} = \xi_{odd}(p,z)$ and $\psi_{odd}=\psi_{odd}(p,z)$, and to
$xi_{even}=\xi_{even}(p,z)$ and $\psi_{even}=\psi_{even}(p,z)$, respectively.
Then, in accord with eq. \ref{ggg},  we find that the latter obey:
\begin{eqnarray}\label{gggg1}
\psi_{odd} &=&1+z  \xi_{even}^2\bigg(\psi_{even}-p\bigg)^2 \,,\\
\psi_{even}&=&1+z  \xi_{odd}^2\bigg(\psi_{odd}-p\bigg)^2 \,,\\
\xi^{-1}_{odd}&=&1+p z \xi_{even}^2+\frac{z \xi_{even}^2 \bigg(\psi_{even}-p\bigg)^2}{1-p}\,, \\
\xi^{-1}_{even}&=&1+p z \xi_{odd}^2+\frac{z \xi_{odd}^2 \bigg(\psi_{odd}-p\bigg)^2}{1-p} \,.\label{gggg4}
\end{eqnarray}
Observe next that
\begin{eqnarray}
\psi_{odd}-p&=&(1-p)\left(\frac{1}{\xi_{odd}}-p z\xi_{even}^2\right)\label{psp} \,, \\
\psi_{even}-p&=&(1-p)\left(\frac{1}{\xi_{even}}-p z \xi_{odd}^2\right) \,,\label{psm}
\end{eqnarray}
such that $\xi_{odd}$  and  $\xi_{even}$ obey, respectively,
\begin{eqnarray}\label{xpp}
 \xi_{odd} \Big(1+p z \xi_{even}^2+{z (1-p)\bigg(1-p z\xi_{odd}^2 \xi_{even}\bigg)^2}\Big)&=&1 \,,\\
 \xi_{even} \Big(1+p z \xi_{odd}^2 +{z (1-p)\bigg(1-p z\xi_{even}^2 \xi_{odd}\bigg)^2}\Big)&=&1 \,.\label{xmm}
\end{eqnarray}
Define next two auxiliary symmetric functions $s=s(p,z)=\xi_{odd}+\xi_{even}$  and $\sigma= \sigma(p,z)=\xi_{odd}- \xi_{even}$. Multiplying both sides of eq. \ref{xpp} by $\xi_{even}$ and of eq.  \ref{xmm} - by $\xi_{odd}$, we arrive,  after some slight rearrangements of both equations, to the following equations which define $s$ and $\sigma$:
\begin{eqnarray}\label{neweq1}
&&\Big(\xi_{odd}-\xi_{even}\Big)\Big(1 + p z \sigma \bigg((1 - p) z \sigma \left(p s z \sigma - 2\right)-s\bigg)\Big)=0 \,,
\end{eqnarray}
and
\begin{eqnarray}\label{neweq2}\nonumber
\fl \Big(\xi_{odd}-\xi_{even}\Big)
\Big((1-p) p^2 \sigma^2 z^3 \left(s^2{-}\sigma \right){-}2 (1-p) p s \sigma  z^2{-}z (p \sigma-1 {+}p){+}1\Big){=}0 \,.
\end{eqnarray}
From eq. \ref{neweq1} we readily find that
\begin{equation}
s =\frac{1-2 (1-p) p z^2 \sigma^2}{p \sigma  z \bigg(1-(1-p) p \sigma^2 z^2\bigg)} \,.
\end{equation}
Substituting this expression  into eq. \ref{neweq2}, we obtain the following expression for $\sigma(p,z)$:
\begin{eqnarray}
 f_{\sigma}(\sigma,p,z)&{=}&- p\sigma z \Bigg[(1-p)\sigma z \bigg(p \sigma z \Big(\sigma z\big((1-p)^2 p^2 \sigma^3 z^3{+}\nonumber\\ \label{sigma}  \fl&&
(1-p) z (p
   (\sigma{-}1){+}1){-}1+p\big){-}1\Big){+}2\bigg){-}1\Bigg]{+}1{=}0.
\end{eqnarray}
Further on,  $\xi_{odd}$ and $\xi_{even}$ are obtained from
\begin{eqnarray}\label{xpxm}
\xi_{odd}=\frac{s+{\sqrt {s^2-4\sigma}}}{2} \,, \xi_{even}=\frac{s-{\sqrt {s^2-4\sigma}}}{2} \,.
\end{eqnarray}
Real values of $\xi_{odd}$ and $\xi_{even}$ obtained in this way reproduce the alternating limits solution, shown by the blue dashed curves in Figs. 7 and  8.

Further on, the  density $\rho_{in}$ within the critical region
can also be expressed through the symmetric auxiliary functions $s$  and $\sigma$.
Substituting next the resulting expression into eq. \ref{kappa_def}, we arrive at our eq. \ref{QQ}, which defines the compressibility within the critical region and is
presented in the main text. In this equation the functions $A_1 = A_1(s,\sigma)$ and $A_2=A_2(s,\sigma)$ are explicitly given by
\begin{equation}\label{a11}
\fl A_1=z\Big((1-p)\left((1-p z\sigma s)^3-6p^2 z^2 \sigma^3+3 p^3 z^3 \sigma^4 s\right)+p(s^3-3\sigma s)\Big),
\end{equation}
and
\begin{eqnarray}\label{a22}
\fl A_2&=&z^2\Bigg((1{-}p)^2(1-pz\sigma s+p^2z^2\sigma^3)^3+
p(1{-}p)\Big(s^3-3\sigma s-3 p z\sigma(s^4-4\sigma s^2+2\sigma^2){+}\nonumber\\ \fl &+&\
3(p z\sigma)^2(s^5-5\sigma s^3+5\sigma^2 s){-}(p z\sigma)^3(s^6-6\sigma s^4+9\sigma^2 s^2+2\sigma^3)
\Big)-p^2\sigma^3\Bigg).
\end{eqnarray}
In turn, the derivatives of $A_1$ and $A_2$ with respect to $z$ obey
\begin{eqnarray}\label{da11}
\fl z A'_1&=&A_1+z^2\Bigg(3(1-p) \Big(-p (1-pz\sigma s)^2(\sigma s+z \left(s \sigma'+\sigma s'\right))-
2p^2z\sigma^2(2\sigma+3 z \sigma')+\nonumber\\
\fl &+&p^3 z^2 \sigma^3(3\sigma s +z \left(4s \sigma'+\sigma s'\right))\Big)+ 3p(s^2 s'-\sigma s'-s \sigma')
\Bigg) \,,
\end{eqnarray}
and
\begin{eqnarray}\label{da22}
\fl z A'_2 &=&2A_2+z^3\Bigg[3(1-p)^2(1-p z \sigma s+p^2 z^2\sigma^3)^2\bigg(-p\Big(\sigma s +z\left(s \sigma'+\sigma s'\right)\Big )+
\nonumber\\
\fl&+&
3p^2z
\Big(2\sigma(\sigma{+}z \sigma')(s^5{-} 5 \sigma s^3{+}5 \sigma^2s){+}5 z\sigma^2(s^4 s'{-}s^3 \sigma'{-}3 s^2 \sigma s'{+}2s\sigma \sigma'{+}\sigma^2 s')\Big){-}
 \nonumber\\
 \fl&-&
3p
\Big((\sigma+z \sigma')(s^4-4 \sigma s^2+2\sigma^2)+4 z \sigma(s^3 s' -s^2 \sigma'-2s \sigma s'+\sigma \sigma')\Big) +
 \nonumber\\
 \fl &+ &
p^2 z \sigma^2(2\sigma +3 z \sigma')\bigg)+3p^2 \sigma^2 \sigma'+p(1-p)\bigg(3 s^2 s'-3\left(s \sigma'+\sigma  s' \right)+
\nonumber\\
\fl &+&
6 z\sigma^3(s^5 s'-s^4 \sigma'-4s^3 \sigma s'+3s\sigma(s \sigma'+\sigma s')+\sigma^2 \sigma') - \nonumber\\
  \fl &-&
p^3 z^2
\Big(3\sigma^2(\sigma+z  \sigma')(s^6-6 s^4\sigma+9\sigma^2 s^2+2\sigma^3)\Big)\bigg)
\Bigg]
\end{eqnarray}
Lastly, we present explicit
expressions for the derivatives $s'$ and  $\sigma'$, which enter the expressions for the
 compressibility :
\begin{eqnarray}\label{derivs}
s' &=&\left(\sigma{+} z \sigma'\right)\Bigg(\frac{1}{pz^2\sigma^2}{-}
\frac{(1{-}p)}{1{-}(1{-}p)p z^2\sigma^2}\left(1{-}\frac{2(1{-}p)p z^2\sigma^2}{1{-}(1{-}p)p z^2\sigma^2}\right)\Bigg)
\end{eqnarray}
and
\begin{equation}\label{derivsigma}
\sigma'=-\frac{\partial f_{\sigma}(\sigma,p,z)}{\partial z} \left(\frac{\partial f_{\sigma}(\sigma,p,z)}{\partial \sigma}\right)^{-1} \,.
\end{equation}

\end{document}